\documentclass[twocolumn]{aastex62}

\usepackage{amsmath,amssymb,amsfonts,mathrsfs}
\usepackage{graphicx}
\usepackage{booktabs,array,multirow}
\usepackage{xcolor}
\usepackage{url}
\usepackage{threeparttable}
\usepackage{xparse}
\usepackage{aas_macros}

\graphicspath{{./}{png/}}

\let\origincludegraphics\includegraphics
\RenewDocumentCommand{\includegraphics}{O{}m}{%
  \IfFileExists{#2}{\origincludegraphics[#1]{#2}}{%
    \fbox{\parbox{0.90\linewidth}{\centering Missing figure: \texttt{\detokenize{#2}}}}%
  }%
}

\begin{document}

\title{A Laplace-Transform Green's Function for Two-Zone Diffusion with Energy Losses: Application to the Geminga Pulsar Halo}

\author{Furong Li}
\affiliation{School of Space and Earth Sciences, Beihang University, Beijing 100191, China}
\email{LiFurong@buaa.edu.cn}

\author{Wei Liu}
\affiliation{State Key Laboratory of Particle Astrophysics, Institute of High Energy Physics, Chinese Academy of Sciences, Beijing 100049, China}
\email{liuwei@ihep.ac.cn}

\author{Yali Shao}
\affiliation{School of Space and Earth Sciences, Beihang University, Beijing 100191, China}
\email{yshao@buaa.edu.cn}

\author{Yi-Qing Guo}
\affiliation{State Key Laboratory of Particle Astrophysics, Institute of High Energy Physics, Chinese Academy of Sciences, Beijing 100049, China}
\affiliation{University of Chinese Academy of Sciences, Beijing 100049, China}
\email{guoyq@ihep.ac.cn}

\begin{abstract}
Extended TeV gamma-ray halos around pulsars are commonly interpreted as evidence for slow diffusion of high-energy electrons and positrons near their sources, with diffusion coefficients about two orders of magnitude below the Galactic average inferred from B/C measurements.
A two-zone diffusion picture offers a natural phenomenological resolution: particles are temporarily confined in a slow-diffusion region around the source and subsequently propagate through the outer region with a diffusion coefficient close to the Galactic average.
Using a Laplace-transform method, we derive an analytical Green’s-function solution to the spherically symmetric two-zone diffusion equation, in which the diffusion coefficient changes discontinuously at $r_b$.
The solution satisfies continuity of both the particle density and the diffusive flux at the boundary and reduces to the standard Gaussian kernel in the one-zone limit. 
For the parameter set used in our comparison, the spatial kernel and the positron flux at Earth agree numerically with an independent integral solution of the same two-zone problem.
Applied to Geminga, the two-zone models examined here produce a more extended outer particle distribution and broader positron spectra than the homogeneous diffusion model, with higher positron fluxes in the TeV range. The adopted injection model further affects the height and shape of the high-energy spectral peak. The predicted gamma-ray surface-brightness profile also reproduces the main radial decline observed by HAWC.

\end{abstract}


\section{Introduction}
\label{sec:introduction}
Recent high-precision measurements of cosmic rays(CRs) have significantly improved our understanding of the origin and propagation of high-energy Galactic leptons.
PAMELA first reported that the CR positron fraction rises above approximately $10\,\mathrm{GeV}$\citep{2009Natur.458..607A, PhysRevLett.105.121101}. This trend was subsequently confirmed by Fermi-LAT, and later measured with much higher precision by AMS-02 up to several hundred GeV.\citep{PhysRevLett.108.011103,PhysRevLett.110.141102,PhysRevLett.113.121101}.
This observed rise deviates from the predictions of conventional CR propagation models, which attribute high-energy positrons predominantly to secondary production\citep{1998ApJ...493..694M}. This discrepancy is commonly referred to as the CR positron excess.
This observation suggests the existence of additional Galactic sources of high-energy electrons and positrons beyond the standard secondary production scenario, while their origin and propagation remain open questions.\citep{1998ApJ...493..694M, 2012APh....39....2S, PhysRevLett.122.041102}.
High-energy electrons and positrons undergo diffusion and energy losses, primarily through synchrotron radiation and inverse-Compton scattering, during their propagation through the Galaxy\citep{2007ARNPS..57..285S, Delahaye2010}. These energy losses limit their propagation distances. As a result, the local high-energy lepton spectra reflect both the injection properties of nearby sources and the characteristics of Galactic CR propagation\citep{PhysRevD.96.023006,2020ChPhC..44h5102T}.

The positron excess has been interpreted mainly in terms of astrophysical sources, particularly pulsars, while more exotic scenarios involving dark-matter annihilation or decay have also been considered\citep{PhysRevD.52.3265, 2009JCAP...01..025H,PhysRevD.88.023001}.
Dark-matter interpretations are constrained by gamma-ray, antiproton, and cosmic microwave background (CMB) observations, with the constraints depending on the annihilation or decay channel\citep{PhysRevLett.115.231301, PhysRevLett.118.191102}. 
Nearby middle-aged pulsars and their pulsar wind nebulae (PWNe) are capable of producing and accelerating large numbers of electron--positron pairs, which are released into the interstellar medium throughout their evolution \citep{PhysRevD.52.3265,2009JCAP...01..025H}. 
Geminga and PSR B0656+14, the pulsar associated with the Monogem supernova remnant, are located within a few hundred parsecs of the Sun and have ages of order $10^5\,\mathrm{yr}$, making them plausible contributors to the local high-energy lepton spectrum \citep{2009JCAP...01..025H,PhysRevD.88.023001,2017Sci...358..911A,2024Symm...16..236L}. 
However, the positron flux reaching Earth depends not only on the injection spectrum, the total energy injected into pairs, and the release history, but also on how the particles propagate near the source and through the surrounding interstellar medium \citep{2017Sci...358..911A,2018ApJ...863...30F,2019MNRAS.484.3491T}.

The discovery of TeV gamma-ray halos has provided an important probe of high-energy particle transport around pulsars. 
The HAWC Collaboration detected spatially extended TeV gamma-ray emission around Geminga and PSR B0656+14 (Monogem)\citep{2017Sci...358..911A}.
Assuming that the emission is produced predominantly through inverse-Compton scattering of photons in the interstellar radiation field (ISRF) by escaped electrons and positrons, its morphology reflects the transport of high-energy leptons around the source.
By fitting the radial surface-brightness profiles of the TeV halos, the HAWC Collaboration inferred diffusion coefficients around Geminga and Monogem that are roughly two to three orders of magnitude below the Galactic average inferred from secondary-to-primary ratios such as B/C \citep{2017Sci...358..911A, 2024ApJ...974..246A}. 
These results point to spatially non-uniform CR transport and pose a challenge to homogeneous diffusion models attempting to account simultaneously for the TeV-halo morphology and the positron flux measured by AMS-02\citep{2018ApJ...863...30F, 2019MNRAS.484.3491T,2018ApJ...869..176L}.
More recently, LHAASO detected extended very-high-energy gamma-ray emission around PSR J0622+3749, providing further observational support for the existence of suppressed-diffusion regions around pulsars\citep{PhysRevLett.126.241103}.

In a two-zone diffusion model, a finite region of suppressed diffusion surrounds the source, while particles diffuse faster outside it \citep{2017Sci...358..911A,2018ApJ...863...30F}. 
The model thus retains the Galactic diffusion parameters constrained by secondary-to-primary ratios such as B/C, while accounting for the suppressed diffusion near the source inferred from TeV-halo observations\citep{2019ApJ...879...91J}.

Fang et al. numerically solved the two-zone diffusion-loss equation for electrons and positrons from Geminga and showed that confining the suppressed diffusion to a finite region around the pulsar allows Geminga to contribute significantly to the local positron excess while remaining compatible with the HAWC observations\citep{2018ApJ...863...30F}.
Tang and Piran derived an analytical solution for a two-zone diffusion model of electrons and positrons from Geminga and showed that it can simultaneously account for the HAWC surface-brightness profile and the local positron spectrum\citep{2019MNRAS.484.3491T}.
Osipov et al. subsequently revisited the analytical treatment of spherical two-zone propagation with radiative losses and obtained a Green's function that satisfies continuity of both the particle density and the diffusive flux at the boundary \citep{2020JPhCS1697a2009O}.
More recently, Liu et al. derived an eigenmode-series Green's function for the two-zone propagation of high-energy CR nuclei\citep{2026arXiv260619701L}.
Bao et al. presented a related series formulation for electron and positron propagation with synchrotron and inverse-Compton energy losses\citep{2026arXiv260626597B}. 


In this work, we construct the propagator using an alternative Laplace-space approach, solving the transport equation separately in the inner and outer regions and determining the regional solutions from the point-source condition at the origin and the matching conditions at the boundary. 
We examine its spatial behavior and apply it to the electron-positron distribution around Geminga, the TeV gamma-ray surface brightness, and the positron flux at Earth. Section~\ref{sec:model} presents the model and analytical derivation, Section~\ref{sec:results} gives numerical tests and applications, and Section~\ref{sec:discussion} summarizes the results and discusses the limitations of the model.


\section{Two-Zone Diffusion model}
\label{sec:model}

\subsection{Cosmic-ray transport equation}
\label{subsec:transport_equation}

The propagation of high-energy electrons and positrons in the Galaxy is primarily determined by spatial diffusion and continuous radiative energy losses\citep{PhysRevD.52.3265}. For a point source, the particle number density, $N(E,\mathbf{r},t)$ , satisfies the following transport equation

\begin{equation}
\begin{aligned}
\frac{\partial N(E,\mathbf{r},t)}{\partial t}
&-\nabla\!\cdot\!\left[D(E,\mathbf{r})\nabla N(E,\mathbf{r},t)\right]
\\
&-\frac{\partial}{\partial E}\left[b(E)N(E,\mathbf{r},t)\right]
=Q(E,t)\delta^{(3)}(\mathbf{r}).
\end{aligned}
\label{eq:transport}
\end{equation}
where $D(E,\mathbf{r})$ is the diffusion coefficient; $b(E)\equiv-dE/dt>0$ is the continuous energy-loss rate; 
The complete source term is $Q(E,t)\delta^{(3)}(\mathbf r)$, $Q$ is the injection rate per unit energy, while the delta function specifies injection at the origin. 
The injection spectrum and time dependence are described in Section~\ref{subsec:source_term}.

We assume isotropic diffusion in a spherically symmetric medium centered on the source, so that $N=N(E,r,t)$ and $D=D(E,r)$, with
$r=|\mathbf r|$. Equation~\eqref{eq:transport} then becomes
\[
\begin{split}
\frac{\partial N}{\partial t}
&-\frac{1}{r^2}\frac{\partial}{\partial r}
\left[r^2D(E,r)\frac{\partial N}{\partial r}\right]
-\frac{\partial}{\partial E}\left[b(E)N\right]\\
&=Q(E,t)\delta^{(3)}(\mathbf r).
\end{split}
\]

We include synchrotron radiation and inverse-Compton scattering as the dominant radiative losses and write \citep{2005MNRAS.363..954M, RevModPhys.42.237}
\begin{equation}
b(E)\simeq\frac{4}{3}c\sigma_T\gamma^2
\left[\sum_i\frac{U_i}{(1+4\gamma\epsilon_i)^{3/2}}
+\frac{B^2}{8\pi}\right],
\label{eq:edot}
\end{equation}
where $\sigma_{\rm T}$ is the Thomson cross section, $\gamma$ is the lepton Lorentz factor, $c$ is the speed of light, and $B$ is the effective magnetic-field strength in the surrounding interstellar medium. 
The index $i$ labels components of the ISRF, including the CMB, infrared radiation, and optical radiation. $U_i$ is the energy density of the
$i$-th radiation component, and $\epsilon_i$ is the corresponding normalized photon energy. For a blackbody component of temperature $T_i$, we adopt
$\epsilon_i=2.8k_BT_i/(m_ec^2)$,
where $k_{\rm B}$ is the Boltzmann constant and $m_e$ is the electron mass\citep{2005MNRAS.363..954M}.
The magnetic and radiation fields are taken to be spatially uniform in both diffusion zones, so that the same loss function $b(E)$ applies throughout the model. Their numerical values are specified in Section~\ref{subsec:density_profiles}.          


Integrating $dE/dt=-b(E)$ gives the time required for a particle to cool from its injection energy $E_0$ to $E$:
\begin{equation}
\tau(E,E_0)
=
\int_E^{E_0}
\frac{dE'}{b(E')} .
\label{eq:cooling_time}
\end{equation}

We further assume that the diffusion coefficient can be factorized into an energy-dependent factor and a spatially dependent factor. In the two-zone model, it is written as
\begin{equation}
D(E,r)
=
\left(\frac{E}{E_\ast}\right)^{\delta}
\begin{cases}
D_{1,0}, & 0<r<r_b,\\[4pt]
D_{2,0}, & r\ge r_b,
\end{cases}
\label{eq:diffusion_coefficient}
\end{equation}
where $r_b$ is the radius of the slow-diffusion region and $D_{1,0}$ and $D_{2,0}$ are the diffusion coefficients in the inner and outer zones at the reference energy $E_\ast$. 
The $D_{1,0}<D_{2,0}$ describes suppressed diffusion near the pulsar and faster transport in the surrounding interstellar medium. 

Diffusion acts while particles cool. Its cooling-dependent diffusion length in zone $i$ is therefore obtained by integrating $D_i$ along the same
energy-loss trajectory
\begin{equation}
r_{d,i}(E,E_0)
=
2
\left[
\int_E^{E_0}
\frac{D_i(E')}{b(E')}dE'
\right]^{1/2},
\qquad i=1,2 .
\label{eq:rd_def}
\end{equation}
Equivalently, defining
\begin{equation}
\Lambda_i(E,E_0)
=
\int_E^{E_0}
\frac{D_i(E')}{b(E')}dE',
\label{eq:Lambda_i}
\end{equation}
the diffusion length can be written as
\begin{equation}
r_{d,i}(E,E_0)
=
2\sqrt{\Lambda_i(E,E_0)} ,
\qquad i=1,2 .
\label{eq:rd_i_def}
\end{equation}


\subsection{Source term}
\label{subsec:source_term}
We model continuous pair injection using the pulsar spin-down luminosity.
For magnetic-dipole braking with index $n=3$, a pulsar of present age $t_{\rm age}$ and initial spin-down timescale $\tau_0$, the spin-down luminosity is then given by\citep{PhysRevD.52.3265}
\begin{equation}
L(t)
=
\dot{E}_{\rm spin}
\left(
1+\frac{t_{\rm age}}{\tau_0}
\right)^2
\left(
1+\frac{t}{\tau_0}
\right)^{-2},
\label{eq:spin_down_luminosity}
\end{equation}
where $t$ is the time elapsed since the birth of the pulsar and $\dot{E}_{\rm spin}$ is its present spin-down power.
\begin{equation}
\begin{aligned}
\dot{E}_{\rm spin}
&=4\pi^2I\frac{\dot P}{P^3}
\\
&\approx3.4\times10^{34}\,\mathrm{erg\,s^{-1}}
\left(\frac{I}{10^{45}\,\mathrm{g\,cm^2}}\right)
\\
&\quad\times
\left(\frac{\dot P}{11.4\times10^{-15}\,\mathrm{s\,s^{-1}}}\right)
\left(\frac{0.237\,\mathrm{s}}{P}\right)^3.
\end{aligned}
\end{equation}

Under the same braking assumption, the pulsar age and the initial spin-down timescale are given by

\begin{equation}
t_{\rm age}
=
\tau_c
\left[
1-\left(\frac{P_0}{P}\right)^2
\right],
\qquad
\tau_0
=
\tau_c
\left(\frac{P_0}{P}\right)^2,
\label{eq:age_tau0}
\end{equation}
where $P_0$ is the initial spin period and $\tau_c=P/(2\dot P)$ is the characteristic age.

We consider either a single power law or a broken power law for the injection spectrum. The broken-power-law form is
\begin{equation}
Q(E,t)
=
Q_0(t)
\begin{cases}
\left(\dfrac{E}{E_b}\right)^{-\alpha_1},
& E_{\min}<E<E_b,\\[6pt]
\left(\dfrac{E}{E_b}\right)^{-\alpha_2},
& E_b\le E<E_{\max},
\end{cases}
\label{eq:injection_broken}
\end{equation}
where $E_b$ denotes the break energy, while $\alpha_1$ and $\alpha_2$ are the spectral indices below and above $E_b$, respectively. For a single power law, one may set $\alpha_1=\alpha_2$ or write
\begin{equation}
Q(E,t)=Q_0(t)E^{-\alpha},
\qquad
E_{\min}<E<E_{\max}.
\label{eq:injection_single}
\end{equation}

The normalization $Q_0(t)$ is determined by the power injected into electron–positron pairs. The parameter $\eta$ denotes the pulsar spin-down luminosity converted into electron–positron pairs. The normalization condition is then
\begin{equation}
\eta L(t)
=
2\int_{E_{\min}}^{E_{\max}}E Q(E,t)dE.
\label{eq:eta_norm}
\end{equation}

\subsection{Two-zone spatial diffusion kernel}
\label{subsec:two_zone_green}

As described in Section~\ref{subsec:transport_equation}, the diffusion coefficients in the
inner and outer regions are assumed to have the same energy
dependence. The diffusion coefficient can therefore be factorized as
\begin{equation}
D(E,r)=\kappa(E)D_0(r),
\label{eq:D_factorization}
\end{equation}
where $\kappa(E)=(E/E_*)^\delta$ and
\begin{equation}
D_0(r)
=
D_{1,0}
+
\left(D_{2,0}-D_{1,0}\right)
\Theta(r-r_b),
\label{eq:D0_heaviside}
\end{equation}
or, equivalently,
\begin{equation}
D_0(r)=
\begin{cases}
D_{1,0}, & 0<r<r_b,\\[4pt]
D_{2,0}, & r>r_b.
\end{cases}
\label{eq:piecewise_D0}
\end{equation}

Under the above assumptions, we first solve the two-zone diffusion
problem in the radial coordinate $r$ and the diffusion-time variable
$t$. 
The Green's function $\mathcal{H}(r,t)$ for a unit instantaneous point source satisfies
\begin{equation}
\frac{\partial \mathcal{H}}{\partial t}
-
\frac{1}{r^2}
\frac{\partial}{\partial r}
\left[
r^2D_0(r)
\frac{\partial \mathcal{H}}{\partial r}
\right]
=
\frac{\delta(r)}{4\pi r^2}\delta(t),
\label{eq:two_zone_diffusion_t}
\end{equation}

Introducing the Laplace transform with respect to $t$,
\begin{equation}
\widetilde{\mathcal{H}}(r,s)
=
\int_0^\infty
e^{-st}\mathcal{H}(r,t)\,dt,
\label{eq:H_laplace}
\end{equation}
we obtain
\begin{equation}
s\widetilde{\mathcal{H}}
-
\frac{1}{r^2}
\frac{d}{dr}
\left[
r^2D_0(r)
\frac{d\widetilde{\mathcal{H}}}{dr}
\right]
=
\frac{\delta(r)}{4\pi r^2}.
\label{eq:H_laplace_full}
\end{equation}

Away from the source position and the boundary,
i.e., for $r\neq0,r_b$, Equation~(\ref{eq:H_laplace_full})
reduces to
\begin{equation}
s\widetilde{\mathcal{H}}_i
-
\frac{D_{i,0}}{r^2}
\frac{d}{dr}
\left(
r^2\frac{d\widetilde{\mathcal{H}}_i}{dr}
\right)
=0,
\qquad i=1,2.
\label{eq:H_laplace_region}
\end{equation}

The general solutions of Equation~(\ref{eq:H_laplace_region})
are derived separately in the inner and outer diffusion regions,
$\widetilde{\mathcal{H}}_{\rm in}(r,s)$ and
$\widetilde{\mathcal{H}}_{\rm out}(r,s)$, respectively.
Their explicit derivation is presented in Appendix~.

At the boundary $r=r_b$, both the particle density
and the diffusive flux are continuous. Therefore,
\begin{equation}
\widetilde{\mathcal{H}}_{\rm in}(r_b,s)
=
\widetilde{\mathcal{H}}_{\rm out}(r_b,s),
\label{eq:bc_density_laplace}
\end{equation}
and
\begin{equation}
D_{1,0}
\left.
\frac{d\widetilde{\mathcal{H}}_{\rm in}}{dr}
\right|_{r=r_b}
=
D_{2,0}
\left.
\frac{d\widetilde{\mathcal{H}}_{\rm out}}{dr}
\right|_{r=r_b}.
\label{eq:bc_flux_laplace}
\end{equation}

The resulting Green's function in Laplace space is
\begin{equation}
\widetilde{\mathcal{H}}(r,s)
=
\begin{cases}
\widetilde{\mathcal{H}}_{\mathrm{in}}(r,s),
& 0<r<r_b,
\\[4pt]
\widetilde{\mathcal{H}}_{\mathrm{out}}(r,s),
& r\geq r_b.
\end{cases}
\label{eq:H_laplace_piecewise}
\end{equation}

The inner-region solution, for $0<r<r_b$, is
\begin{equation}
\begin{aligned}
\widetilde{\mathcal{H}}_{\mathrm{in}}(r,s)
={}&
\frac{1}{4\pi rD_{1,0}}
\sum_{n=0}^{\infty}(-1)^n
\Bigg[
\\
&\sum_{m=0}^{n}
\binom{n}{m}
R_{\mathrm{c}}^{\,n-m}R_0^{\,m}
\left(\sqrt{s}+\lambda\right)^{-m}
\\
&\qquad\times
\exp\left[
-\sqrt{s}\,
\frac{2nr_b+r}{\sqrt{D_{1,0}}}
\right]
\\
&+
\sum_{m=0}^{n+1}
\binom{n+1}{m}
R_{\mathrm{c}}^{\,n+1-m}R_0^{\,m}
\left(\sqrt{s}+\lambda\right)^{-m}
\\
&\qquad\times
\exp\left[
-\sqrt{s}\,
\frac{2(n+1)r_b-r}{\sqrt{D_{1,0}}}
\right]
\Bigg].
\end{aligned}
\label{eq:H_laplace_inner_series}
\end{equation}

The outer-region solution, for $r\geq r_b$, is
\begin{equation}
\begin{aligned}
\widetilde{\mathcal{H}}_{\mathrm{out}}(r,s)
={}&
-\frac{2r_b}{4\pi r}
\sum_{n=0}^{\infty}(-1)^n
\Bigg\{
\\
&\frac{1}{
-r_b\left(
D_{1,0}+\sqrt{D_{1,0}D_{2,0}}
\right)}
\\
&\qquad\times
\sum_{m=0}^{n}
\binom{n}{m}
R_{\mathrm{c}}^{\,n-m}R_0^{\,m}
\left(\sqrt{s}+\lambda\right)^{-m}
\\
&-
\frac{\lambda}{
-r_b\left(
D_{1,0}+\sqrt{D_{1,0}D_{2,0}}
\right)}
\\
&\qquad\times
\sum_{m=0}^{n}
\binom{n}{m}
R_{\mathrm{c}}^{\,n-m}R_0^{\,m}
\left(\sqrt{s}+\lambda\right)^{-(m+1)}
\Bigg\}
\\
&\times
\exp\left\{
-\sqrt{s}
\left[
\frac{(2n+1)r_b}{\sqrt{D_{1,0}}}
+
\frac{r-r_b}{\sqrt{D_{2,0}}}
\right]
\right\}.
\end{aligned}
\label{eq:H_laplace_outer_series}
\end{equation}

where $R_{\mathrm{c}}=\frac{
\sqrt{D_{1,0}}-\sqrt{D_{2,0}}
}{
\sqrt{D_{1,0}}+\sqrt{D_{2,0}}
},$
$R_0=\frac{2\left(D_{1,0}-D_{2,0}\right)\sqrt{D_{1,0}}
}{
r_b
\left(
\sqrt{D_{1,0}}+\sqrt{D_{2,0}}
\right)^2
},$
$\lambda=-\frac{D_{1,0}-D_{2,0}}{r_b\left(\sqrt{D_{1,0}}+\sqrt{D_{2,0}}\right)}$

Equations~(\ref{eq:H_laplace_inner_series}) and
(\ref{eq:H_laplace_outer_series}) show that each term of the
Laplace-space Green's function has the generic form
\begin{equation}
\frac{
\exp(-a\sqrt{s})
}{
\left(\sqrt{s}+\lambda\right)^m
},
\label{eq:generic_laplace_term}
\end{equation}
The inverse Laplace transformation can therefore be
performed term by term.

For $m=1$, the required inverse Laplace transform is
\begin{equation}
\begin{aligned}
f_1(a,\lambda,t)
&\equiv\mathcal{L}^{-1}
\!\left\{\frac{e^{-a\sqrt{s}}}{\sqrt{s}+\lambda}\right\}
\\
&=\frac{e^{-a^2/(4t)}}{\sqrt{\pi t}}
-\lambda e^{a\lambda+\lambda^2t}
\operatorname{erfc}\!\left(
\frac{a}{2\sqrt{t}}+\lambda\sqrt{t}
\right).
\end{aligned}
\label{eq:inverse_laplace_f1}
\end{equation}

For an integer $m\geq 1$, the higher-order inverse transforms are
obtained from
\begin{equation}
\begin{aligned}
f_m(a,\lambda,t)
&\equiv\mathcal{L}^{-1}
\!\left\{\frac{e^{-a\sqrt{s}}}
{(\sqrt{s}+\lambda)^m}\right\}
=\frac{(-1)^{m-1}}{(m-1)!}
\frac{\partial^{m-1}}{\partial\lambda^{m-1}}
 f_1(a,\lambda,t).
\end{aligned}
\label{eq:inverse_laplace_fm}
\end{equation}

For $m=0$, the corresponding inverse transform is
\begin{equation}
f_0(a,t)
\equiv
\mathcal{L}^{-1}
\left\{
\exp(-a\sqrt{s})
\right\}
=
\frac{a}{2\sqrt{\pi}\,t^{3/2}}
\exp\left(
-\frac{a^2}{4t}
\right).
\label{eq:inverse_laplace_f0}
\end{equation}

Combining the above inverse Laplace transforms with the energy relation given in Section~\ref{subsec:transport_equation}, we obtain

\begin{equation}
H(r;E,E_0)
=
\begin{cases}
H_{\rm in}(r;E,E_0), & 0<r<r_b,\\[4pt]
H_{\rm out}(r;E,E_0), & r\ge r_b,
\end{cases}
\label{eq:H_piecewise}
\end{equation}

For the inner region $0<r<r_b$, the two-zone spatial diffusion kernel is written as
\begin{widetext}
\begin{align}
H_{\rm in}(r;E,E_0)
&=
\frac{1}{\pi r r_{d,1}^2}
\sum_{n=0}^{\infty}
(-1)^n
\Bigg\{
\nonumber\\
&\quad
\left(
\frac{r_{d,1}-r_{d,2}}
{r_{d,1}+r_{d,2}}
\right)^n
\frac{2nr_b+r}{\sqrt{\pi}r_{d,1}}
\exp\left[
-\frac{(2nr_b+r)^2}{r_{d,1}^2}
\right]
\nonumber\\
&\quad
+
\left(
\frac{r_{d,1}-r_{d,2}}
{r_{d,1}+r_{d,2}}
\right)^{n+1}
\frac{2(n+1)r_b-r}{\sqrt{\pi}r_{d,1}}
\exp\left[
-\frac{\left(2(n+1)r_b-r\right)^2}{r_{d,1}^2}
\right]
\nonumber\\
&\quad
+
\sum_{m=1}^{n}
\binom{n}{m}
\left(
\frac{r_{d,1}-r_{d,2}}
{r_{d,1}+r_{d,2}}
\right)^{n-m}
\left[
\frac{r_{d,1}(r_{d,1}-r_{d,2})}
{r_b(r_{d,1}+r_{d,2})}
\right]^m
(-1)^{m-1}
\frac{1}{(m-1)!}
\nonumber\\
&\qquad\qquad
\times
\frac{\partial^{m-1}}{\partial q^{m-1}}
\Bigg[
\frac{1}{\sqrt{\pi}}
\exp\left(
-\frac{(2nr_b+r)^2}{r_{d,1}^2}
\right)
\nonumber\\
&\qquad\qquad\qquad\qquad
-
q
\exp\left(
\frac{2(2nr_b+r)q}{r_{d,1}}+q^2
\right)
\operatorname{erfc}
\left(
\frac{2nr_b+r}{r_{d,1}}+q
\right)
\Bigg]
\nonumber\\
&\quad
+
\sum_{m=1}^{n+1}
\binom{n+1}{m}
\left(
\frac{r_{d,1}-r_{d,2}}
{r_{d,1}+r_{d,2}}
\right)^{n+1-m}
\left[
\frac{r_{d,1}(r_{d,1}-r_{d,2})}
{r_b(r_{d,1}+r_{d,2})}
\right]^m
(-1)^{m-1}
\frac{1}{(m-1)!}
\nonumber\\
&\qquad\qquad
\times
\frac{\partial^{m-1}}{\partial q^{m-1}}
\Bigg[
\frac{1}{\sqrt{\pi}}
\exp\left(
-\frac{\left(2(n+1)r_b-r\right)^2}{r_{d,1}^2}
\right)
\nonumber\\
&\qquad\qquad\qquad\qquad
-
q
\exp\left(
\frac{2\left(2(n+1)r_b-r\right)q}{r_{d,1}}+q^2
\right)
\operatorname{erfc}
\left(
\frac{2(n+1)r_b-r}{r_{d,1}}+q
\right)
\Bigg]
\Bigg\}.
\label{eq:H_inner_expanded_q}
\end{align}
\end{widetext}

For the outer region $r\ge r_b$, the spatial diffusion kernel is
\begin{widetext}
\begin{align}
H_{\rm out}(r;E,E_0)
&=
-\frac{2r_b}{4\pi r}
\sum_{n=0}^{\infty}
(-1)^n
\Bigg\{
\nonumber\\
&\quad
-\frac{4}
{r_b r_{d,1}(r_{d,1}+r_{d,2})}
\Bigg[
\nonumber\\
&\qquad
\left(
\frac{r_{d,1}-r_{d,2}}
{r_{d,1}+r_{d,2}}
\right)^n
\frac{1}{2\sqrt{\pi}}
\left[
\frac{2(2n+1)r_b}{r_{d,1}}
+
\frac{2(r-r_b)}{r_{d,2}}
\right]
\nonumber\\
&\qquad\qquad
\times
\exp\left\{
-\frac{1}{4}
\left[
\frac{2(2n+1)r_b}{r_{d,1}}
+
\frac{2(r-r_b)}{r_{d,2}}
\right]^2
\right\}
\nonumber\\
&\qquad
+
\sum_{m=1}^{n}
\binom{n}{m}
\left(
\frac{r_{d,1}-r_{d,2}}
{r_{d,1}+r_{d,2}}
\right)^{n-m}
\left[
\frac{r_{d,1}(r_{d,1}-r_{d,2})}
{r_b(r_{d,1}+r_{d,2})}
\right]^m
(-1)^{m-1}
\frac{1}{(m-1)!}
\nonumber\\
&\qquad\qquad
\times
\frac{\partial^{m-1}}{\partial q^{m-1}}
\Bigg[
\frac{1}{\sqrt{\pi}}
\exp\left\{
-\frac{1}{4}
\left[
\frac{2(2n+1)r_b}{r_{d,1}}
+
\frac{2(r-r_b)}{r_{d,2}}
\right]^2
\right\}
\nonumber\\
&\qquad\qquad\qquad\qquad
-
q
\exp\left\{
\left[
\frac{2(2n+1)r_b}{r_{d,1}}
+
\frac{2(r-r_b)}{r_{d,2}}
\right]q
+q^2
\right\}
\nonumber\\
&\qquad\qquad\qquad\qquad
\times
\operatorname{erfc}
\left[
\frac{(2n+1)r_b}{r_{d,1}}
+
\frac{r-r_b}{r_{d,2}}
+
q
\right]
\Bigg]
\Bigg]
\nonumber\\
&\quad
+
\frac{4q}
{r_b r_{d,1}(r_{d,1}+r_{d,2})}
\Bigg[
\nonumber\\
&\qquad
\sum_{m=0}^{n}
\binom{n}{m}
\left(
\frac{r_{d,1}-r_{d,2}}
{r_{d,1}+r_{d,2}}
\right)^{n-m}
\left[
\frac{r_{d,1}(r_{d,1}-r_{d,2})}
{r_b(r_{d,1}+r_{d,2})}
\right]^m
(-1)^m
\frac{1}{m!}
\nonumber\\
&\qquad\qquad
\times
\frac{\partial^{m}}{\partial q^{m}}
\Bigg[
\frac{1}{\sqrt{\pi}}
\exp\left\{
-\frac{1}{4}
\left[
\frac{2(2n+1)r_b}{r_{d,1}}
+
\frac{2(r-r_b)}{r_{d,2}}
\right]^2
\right\}
\nonumber\\
&\qquad\qquad\qquad\qquad
-
q
\exp\left\{
\left[
\frac{2(2n+1)r_b}{r_{d,1}}
+
\frac{2(r-r_b)}{r_{d,2}}
\right]q
+q^2
\right\}
\nonumber\\
&\qquad\qquad\qquad\qquad
\times
\operatorname{erfc}
\left[
\frac{(2n+1)r_b}{r_{d,1}}
+
\frac{r-r_b}{r_{d,2}}
+
q
\right]
\Bigg]
\Bigg]
\Bigg\}.
\label{eq:H_outer_expanded_q}
\end{align}
\end{widetext}

where
\[
q=\frac{r_{d,2}-r_{d,1}}{2r_b}.
\]

At the boundary $r=r_b$, equations~(\ref{eq:H_inner_expanded_q}) and
(\ref{eq:H_outer_expanded_q}) satisfy continuity of the particle density and the diffusive flux:
\begin{equation}
H_{\rm in}(r_b;E,E_0)=H_{\rm out}(r_b;E,E_0),
\label{eq:bc_density}
\end{equation}
\begin{equation}
D_1\left.\frac{\partial H_{\rm in}}{\partial r}\right|_{r=r_b}
=
D_2\left.\frac{\partial H_{\rm out}}{\partial r}\right|_{r=r_b}.
\label{eq:bc_flux}
\end{equation}


In the one-zone limit, $D_1=D_2$, such that $r_{d,1}=r_{d,2} \equiv r_d$, both the inner-region and outer-region kernels reduce to the standard three-dimensional diffusion Green's function
\begin{equation}
H_{\rm 1z}(r;r_d)
=
\frac{1}{\pi^{3/2}r_d^3}
\exp\left(
-\frac{r^2}{r_d^2}
\right).
\label{eq:one_zone_kernel_rd}
\end{equation}

\subsection{Continuous injection solution}
\label{subsec:two_zone_N}

Let the age of the source be $t_{\rm age}$. A particle injected at time $t_0$ with energy $E_0$ and cooled to energy $E$  at the present time satisfies
\begin{equation}
t_{\rm age}-t_0=\tau(E,E_0).
\label{eq:cooling_relation}
\end{equation}

For a continuously injecting source of age $t_{\rm age}$, the solution to Equation~(\ref{eq:transport}) gives the particle number density as

\begin{equation}
\begin{aligned}
N(E,r,t)
={}&\int_0^t \!dt_0\int_E^{E_{\max}}\!dE_0\,
Q(E_0,t_0) G(E,E_0,\mathbf{r},t,t_0).
\end{aligned}
\label{eq:N_general}
\end{equation}

Using the cooling relation in Equation ~(\ref{eq:cooling_relation}), the particle number density for continuous injection can be written as  
\begin{align}
N(E,r,t_{\rm age})
&=
\int_{0}^{\min\left[t_{\rm age},\,\tau(E,E_{\max})\right]}
d\Delta t\,
\frac{b\!\left[E_0(E,\Delta t)\right]}{b(E)}
\nonumber\\
&\quad\times
Q\!\left[E_0(E,\Delta t),t_{\rm age}-\Delta t\right]
H\!\left[r;E,E_0(E,\Delta t)\right],
\label{eq:N_Es_form}
\end{align}

where $\Delta t=t_{\rm age}-t_0$, and $E_0(E,\Delta t)$ is the injection energy required for a particle to cool to energy $E$ over the time interval $\Delta t$. The spatial diffusion kernel $H(r;E,E_0)$ is given in Section~\ref{subsec:two_zone_green}.

\subsection{Positron flux at the Earth}
\label{subsec:model_positron_flux}

The flux of positron at Earth is given by\citep{PhysRevD.100.123015}
\begin{equation}
J_{e^+}(E)
\simeq
\frac{c}{4\pi}
N_{e^+}(E,d,t_{\rm age}).
\label{eq:positron_flux}
\end{equation}
where $d$ is the distance from Geminga to the Earth.

\subsection{Inverse-Compton emission and gamma-ray surface brightness}
\label{subsec:model_gamma_brightness}

The propagated electrons and positrons produce gamma rays through
inverse-Compton scattering on the ISRF. The differential photon production rate per lepton is
written as\citep{PhysRev.167.1159, RevModPhys.42.237, PhysRevD.100.123015}
\begin{equation}
\mathcal{P}_{\rm IC}(E_e,E_\gamma)
=
c\int_0^\infty d\varepsilon\,
n_{\rm ph}(\varepsilon)
\frac{d\sigma_{\rm IC}}{dE_\gamma},
\label{eq:ic_single_particle}
\end{equation}
where $\varepsilon$ is the target-photon energy,
$n_{\rm ph}(\varepsilon)$ is the differential photon number density,
and $d\sigma_{\rm IC}/dE_\gamma$ is the Klein--Nishina differential
cross section.

The inverse-Compton spectrum is evaluated using the \texttt{InverseCompton} model implemented in \texttt{naima} \citep{2015ICRC...34..922Z}, including Klein--Nishina effects. 
The inverse-Compton emission and the energy-loss calculation use the same radiation-field components, with the temperatures and energy densities specified in Section~\ref{subsec:density_profiles}.

Assuming equal injection of electrons and positrons,
$N_{e^-}=N_{e^+}$, the differential gamma-ray volume emissivity is
\begin{equation}
q_\gamma(E_\gamma,r)
=
2\int_{m_ec^2}^{\infty} dE_e\,
N_{e^+}(E_e,r,t_{\rm age})
\mathcal{P}_{\rm IC}(E_e,E_\gamma),
\label{eq:gamma_emissivity}
\end{equation}

where the factor of 2 accounts for the equal contributions of
electrons and positrons.

Assuming isotropic IC emission, the differential gamma-ray
intensity at an angular distance $\theta$ from Geminga is
\begin{equation}
I_\gamma(E_\gamma,\theta)
=
\frac{1}{4\pi}
\int_0^\infty ds\,
q_\gamma\!\left[E_\gamma,r(s,\theta)\right],
\label{eq:gamma_intensity}
\end{equation}
where
\begin{equation}
r(s,\theta)
=
\left(
d^2+s^2-2ds\cos\theta
\right)^{1/2}.
\label{eq:los_radius}
\end{equation}
and $d$ is the distance from the Earth to Geminga.

The surface-brightness profile is integrated over the photon-energy range of $5$--$50\,{\rm TeV}$.

\section{Results}
\label{sec:results}

\subsection{Spatial properties of the analytical two-zone kernel}
\label{subsec:kernel_results}

\begin{figure*}[htbp]
\centering

\begin{minipage}{0.48\textwidth}
\centering
\includegraphics[width=\linewidth]{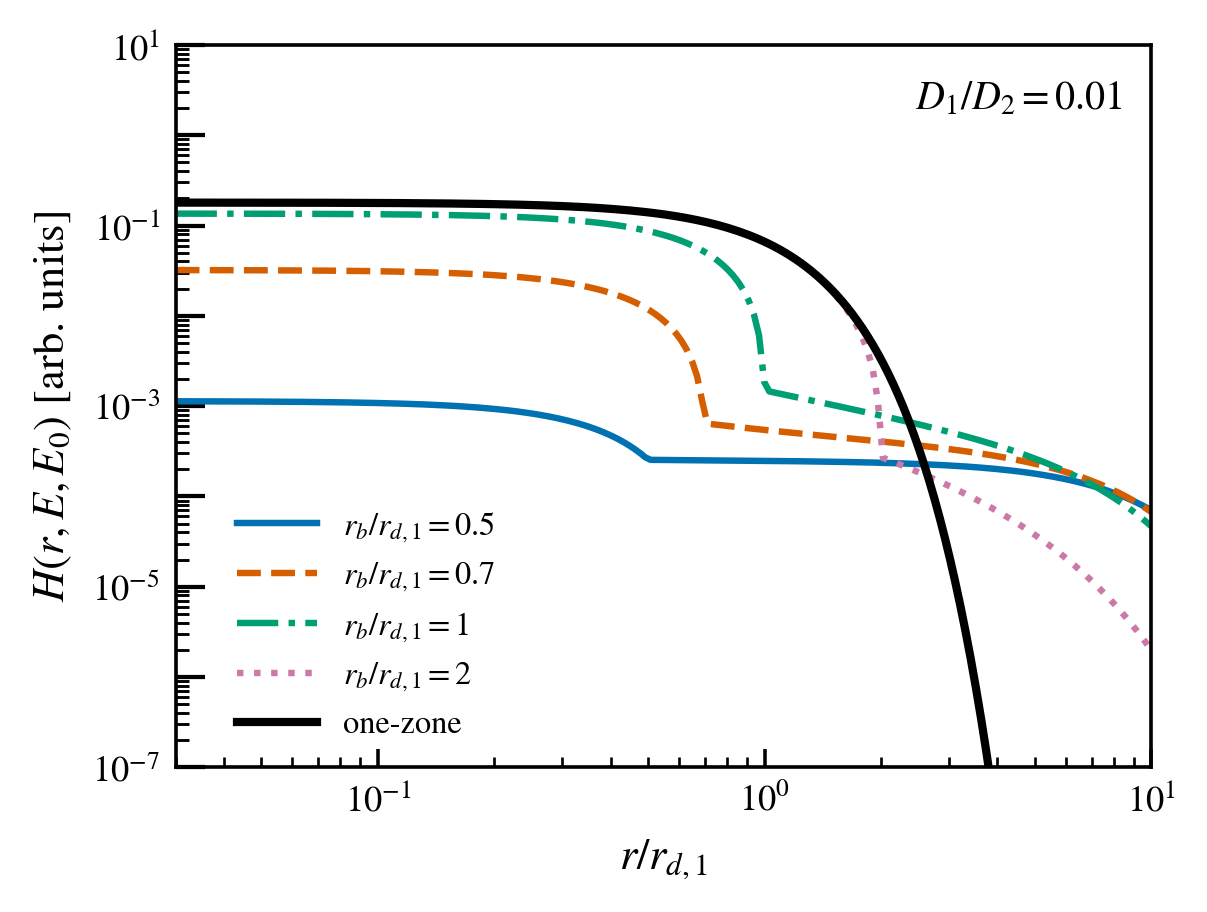}
\\
(a)
\end{minipage}
\hfill
\begin{minipage}{0.48\textwidth}
\centering
\includegraphics[width=\linewidth]{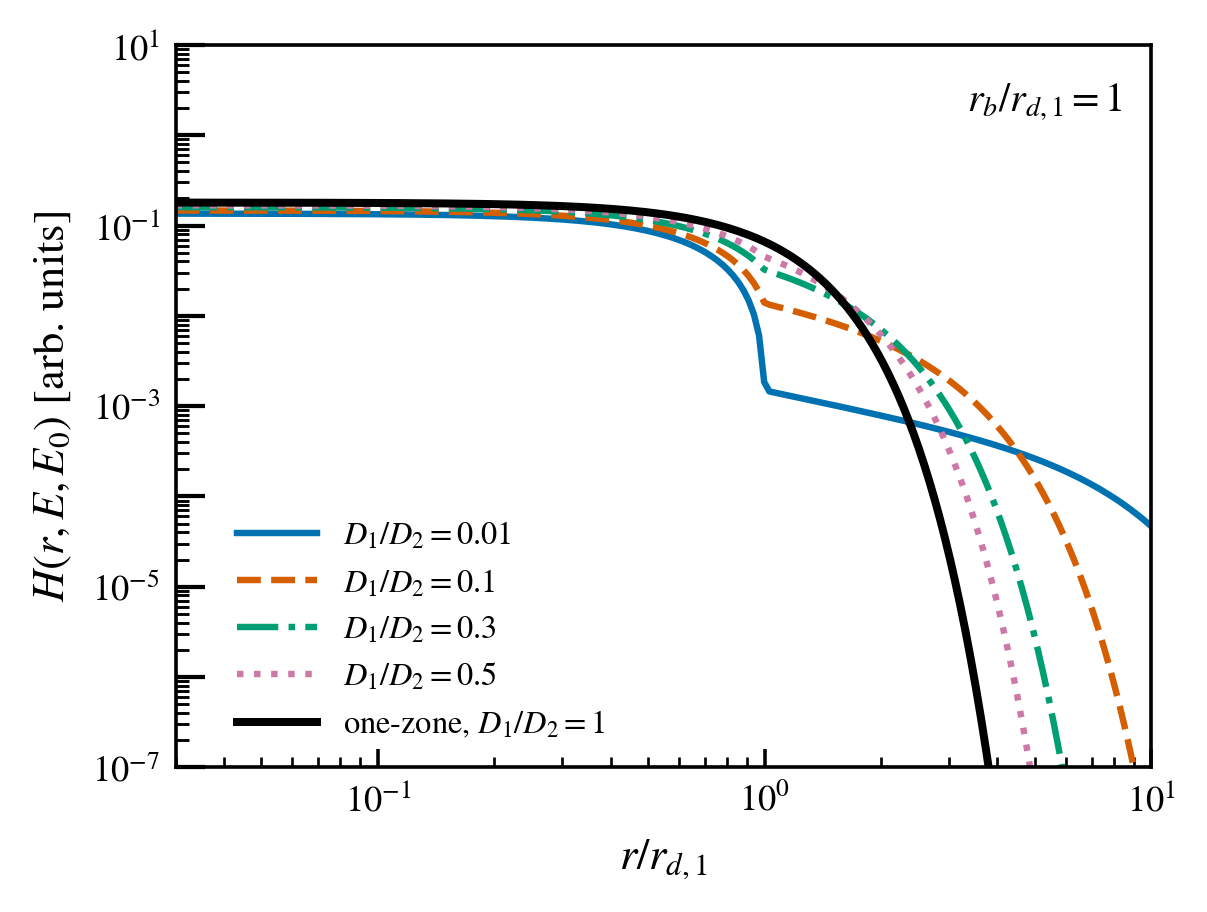}
\\
(b)
\end{minipage}

\caption{The two-zone diffusion kernel $H(r;E,E_0)$. (a) Dependence on the relative boundary scale $r_b/r_{d,1}$ for fixed $D_1/D_2=0.01$. (b) Dependence on the diffusion-coefficient ratio $D_1/D_2$ for fixed $r_b/r_{d,1}=1$. The ordinate is the dimensionless quantity $r_{d,1}^{3}H$. The black solid curve in each panel shows the one-zone result.}

\label{fig:kernel}

\end{figure*}

The two-zone diffusion kernel derived in Section~\ref{subsec:two_zone_green} assumes the same energy dependence of the diffusion coefficient in the inner and outer regions. Its spatial profile is mainly controlled by the dimensionless boundary scale $r_b/r_{d,1}$ and the diffusion-coefficient ratio $D_1/D_2$. Adopting a boundary radius of $r_b=60~\mathrm{pc}$, Figure~\ref{fig:kernel} illustrates the dependence of the radial kernel $H_(r;E,E_0)$ on these two parameters. 
In panel (a), $D_1/D_2=0.01$ is fixed while $r_b/r_{d,1}$ is varied. 
In panel (b), $r_b/r_{d,1}=1$ is fixed and $D_1/D_2$ is varied. 
All two-zone profiles show a clear change in slope near the boundary. 
Smaller values of $D_1/D_2$ produce a broader distribution with a more extended tail. As $D_1/D_2$ increases toward 1, the difference between the two regions decreases and the kernel approaches the one-zone result.

We evaluate the density- and flux-matching residuals for the parameter combinations listed in Table~\ref{tab:kernel_validation} to check the numerical implementation.
We define the density-continuity residual as
\begin{equation}
\epsilon_H
=
\frac{
\left|
H_{\rm in}(r_b)-H_{\rm out}(r_b)
\right|
}{
\max\left[
|H_{\rm in}(r_b)|,
|H_{\rm out}(r_b)|
\right]
},
\end{equation}
and the diffusive-flux-continuity residual as
\begin{equation}
\epsilon_J
=
\frac{
\left|
D_1\partial_rH_{\rm in}(r_b)
-
D_2\partial_rH_{\rm out}(r_b)
\right|
}{
\max\left[
|D_1\partial_rH_{\rm in}(r_b)|,
|D_2\partial_rH_{\rm out}(r_b)|
\right]
}.
\end{equation}

The maximum boundary-matching residuals obtained in the parameter
scan are listed in Table~\ref{tab:kernel_validation}.

\begin{table*}[htbp]
\centering
\caption{Numerical verification of the boundary matching conditions
for the two-zone spatial diffusion kernel.}
\label{tab:kernel_validation}

\begin{threeparttable}

\begin{tabular}{lcc}
\toprule
Validation test
& Definition
& Maximum residual \\
\midrule

Density continuity
& $\epsilon_H$
& $7.2\times10^{-14}$ \\

Diffusive-flux continuity
& $\epsilon_J$
& $8.5\times10^{-7}$ \\

\bottomrule
\end{tabular}

\begin{tablenotes}[flushleft]
\footnotesize

\item \emph{Notes.}
The residuals are the maximum values obtained over
$D_1/D_2=0.01,\ 0.05,\ 0.10,\ 0.50$ and
$r_b/r_{d,1}=0.5,\ 0.7,\ 1.0,\ 2.0$.

\end{tablenotes}

\end{threeparttable}
\end{table*}

\subsection{Radial positron density in the two-zone model}
\label{subsec:density_profiles}

We next examine the spatial distribution of positrons in the two-zone diffusion model,
using the following numerical parameters for Geminga.
We quote diffusion coefficients at $E_*=100\,\mathrm{TeV}$ and adopt $\delta=1/3$\citep{2017Sci...358..911A}.
We fix $D_{1,0}=4.5\times10^{27}\,\mathrm{cm^2\,s^{-1}}$, the central value of the HAWC joint fit to the Geminga and Monogem halos.
We fix $D_{1,0}$ and vary the ratio $D_{1}/D_{2}$.
The propagation parameters are listed in Table~\ref{tab:propagation_parameters} .

We adopt an effective interstellar magnetic field of $B=3\,\mu\mathrm G$ in both zones. The radiation field consists of the CMB, infrared radiation, and optical radiation, with $(T_i,U_i)=(2.7\,\mathrm K,0.26\,\mathrm{eV\,cm^{-3}})$, $(20\,\mathrm K,0.30\,\mathrm{eV\,cm^{-3}})$, and $(5000\,\mathrm K,0.30\,\mathrm{eV\,cm^{-3}})$, respectively.
For the pulsar, we use $P=0.237\,\mathrm s$ and retain the rounded benchmarks $\tau_c\simeq340\,\mathrm{kyr}$ and $\dot E_{\rm spin}\simeq3.3\times10^{34}\,\mathrm{erg\,s^{-1}}$ used in the HAWC analysis \citep{2017Sci...358..911A}.
The adopted rounded $P$ and $\tau_c$ imply $\dot P\simeq1.10\times10^{-14}\,\mathrm{s\,s^{-1}}$.
The injection parameters are summarized in Table~\ref{tab:injection_parameters},
with $E_{\min}=1\,\mathrm{GeV}$ and $E_{\max}=500\,\mathrm{TeV}$.

\begin{figure}[htbp]
\centering
\includegraphics[width=1.0\linewidth]{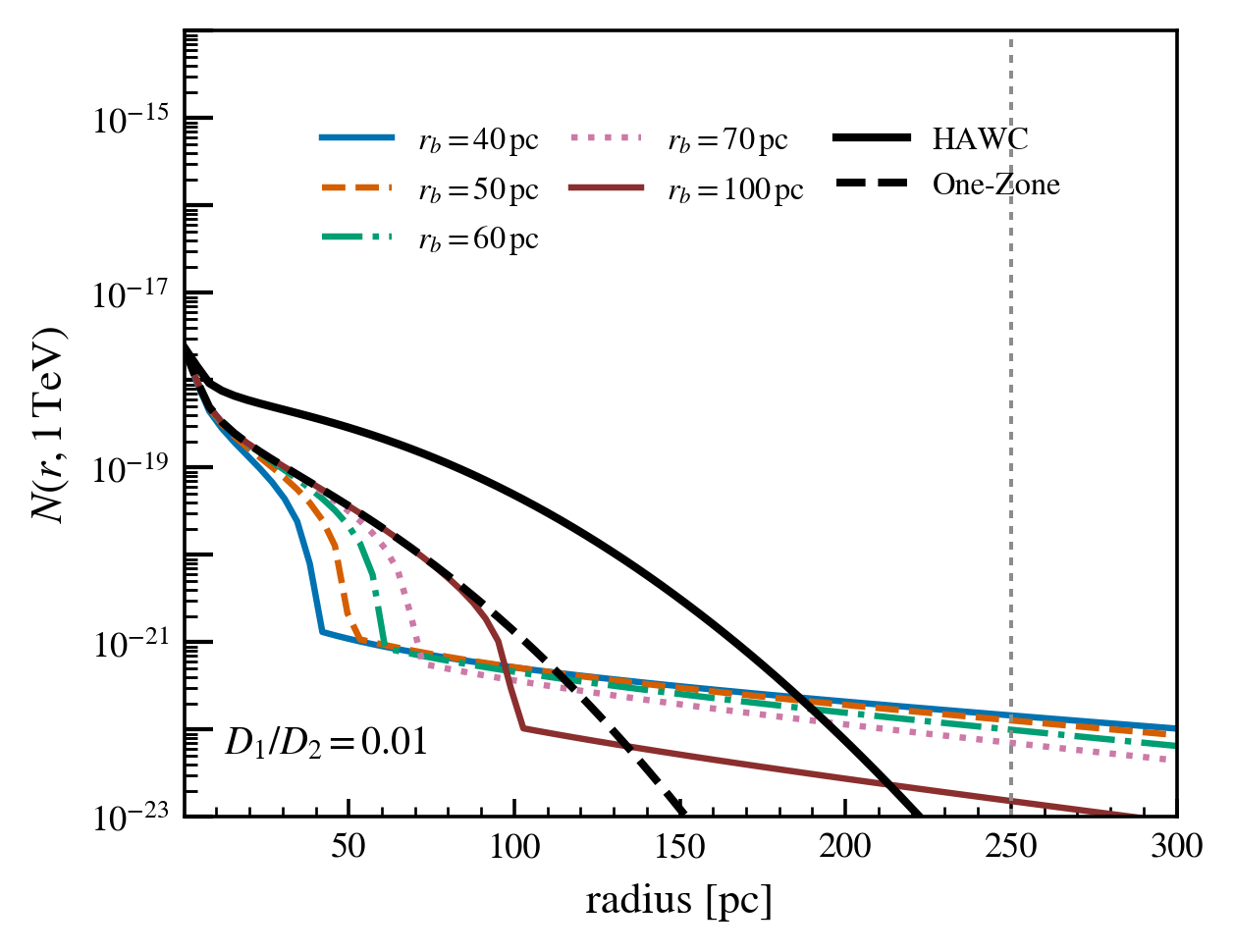}
\caption{$N(r,\,1\,TeV)$ for different values of $r_b$, with $D_1/D_2=0.01$. The gray dashed line marks the distance from Geminga to the Earth, $250\,\mathrm{pc}$ \citep{2012ApJ...755...39V}, while the black dashed line represents the one-zone reference model.
}
\label{fig:density-rb}
\end{figure}

Figure~\ref{fig:density-rb} shows $N(r,1\,\mathrm{TeV})$ for a single-power-law injection spectrum, a fixed diffusion-coefficient ratio $D_1/D_2=0.01$, and different values of the slow-zone radius $r_b$.
As $r_b$ increases, the transition moves toward larger distances. At distances of a few hundred parsecs, $N(r,1\,\mathrm{TeV})$ is higher for smaller $r_b$. For $r_b = 100 \,pc$, the positron density is lower than that for smaller values of $r_b$. At the location of the Earth, approximately $250 \,pc$ from Geminga, all two-zone models give higher positron densities than the one-zone model.

\begin{table*}[htbp]
\centering
\caption{Propagation parameters adopted in this work.}
\label{tab:propagation_parameters}

\begin{threeparttable}
\begin{tabular}{lcccc}
\toprule
Propagation model
& $D_{1,0}$
& $D_{2,0}$
& $D_{1}/D_{2}$
& $r_b$  \\
& $[10^{27}\,\mathrm{cm^2\,s^{-1}}]$
& $[10^{27}\,\mathrm{cm^2\,s^{-1}}]$
& 
& [pc]\\
\midrule

One-zone model
& $4.5$
& $4.5$
& $1$
& -- \\

Two-zone model 
& $4.5$
& $9.0$
& $0.50$
& $60$ \\

Two-zone model 
& $4.5$
& $45$
& $0.10$
& $60$ \\

Two-zone model 
& $4.5$
& $90$
& $0.05$
& $60$ \\

Two-zone model 
& $4.5$
& $450$
& $0.01$
& $60$ \\

\bottomrule
\end{tabular}

\begin{tablenotes}[flushleft]
\footnotesize
\item \emph{Notes.}
Both zones have the same energy dependence,
$D_i(E)=D_{i,0}(E/E_\ast)^\delta$, with $\delta=1/3$.
\end{tablenotes}

\end{threeparttable}
\end{table*}

\begin{table*}[htbp]
\centering
\caption{Injection parameters adopted for Geminga.}
\label{tab:injection_parameters}

\begin{threeparttable}
\begin{tabular}{lccccc}
\toprule
Injection model
& $\alpha_1$
& $\alpha_2$
& $E_b$ 
& $P_0$ 
& $\eta$ \\
&
&
&[TeV]
&[s]
&\\
\midrule

HAWC 
& $2.34$
& $2.34$
& --
& $0.045$
& $0.40$ \\

Single-power-law injection
& $2.34$
& $2.34$
& --
& $0.14$
& $0.40$ \\

Broken-power-law injection
& $1.50$
& $2.34$
& $30$
& $0.045$
& $0.015$ \\

\bottomrule
\end{tabular}

\end{threeparttable}
\end{table*}

\begin{figure*}[htbp]

\centering

\begin{minipage}{0.48\textwidth}
\centering
\includegraphics[width=\linewidth]{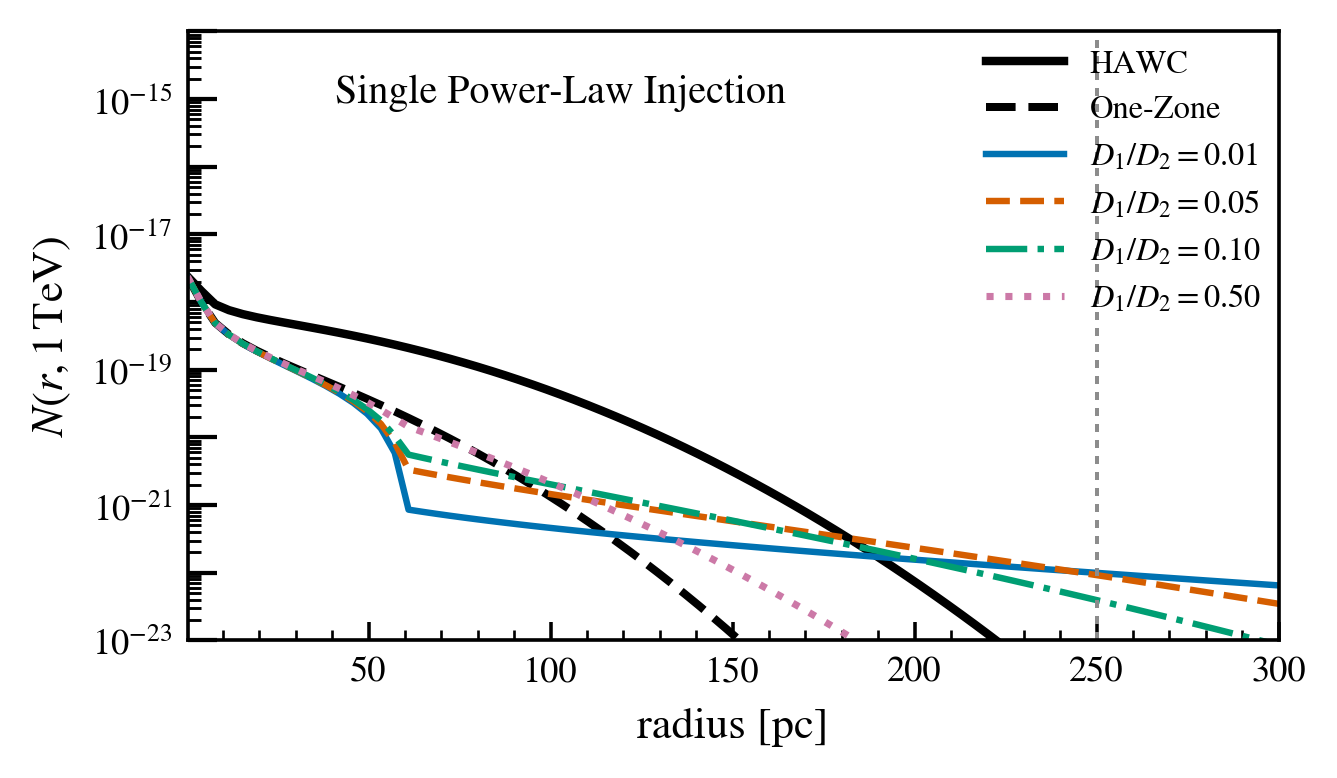}
\\
(a)
\end{minipage}
\hfill
\begin{minipage}{0.48\textwidth}
\centering
\includegraphics[width=\linewidth]{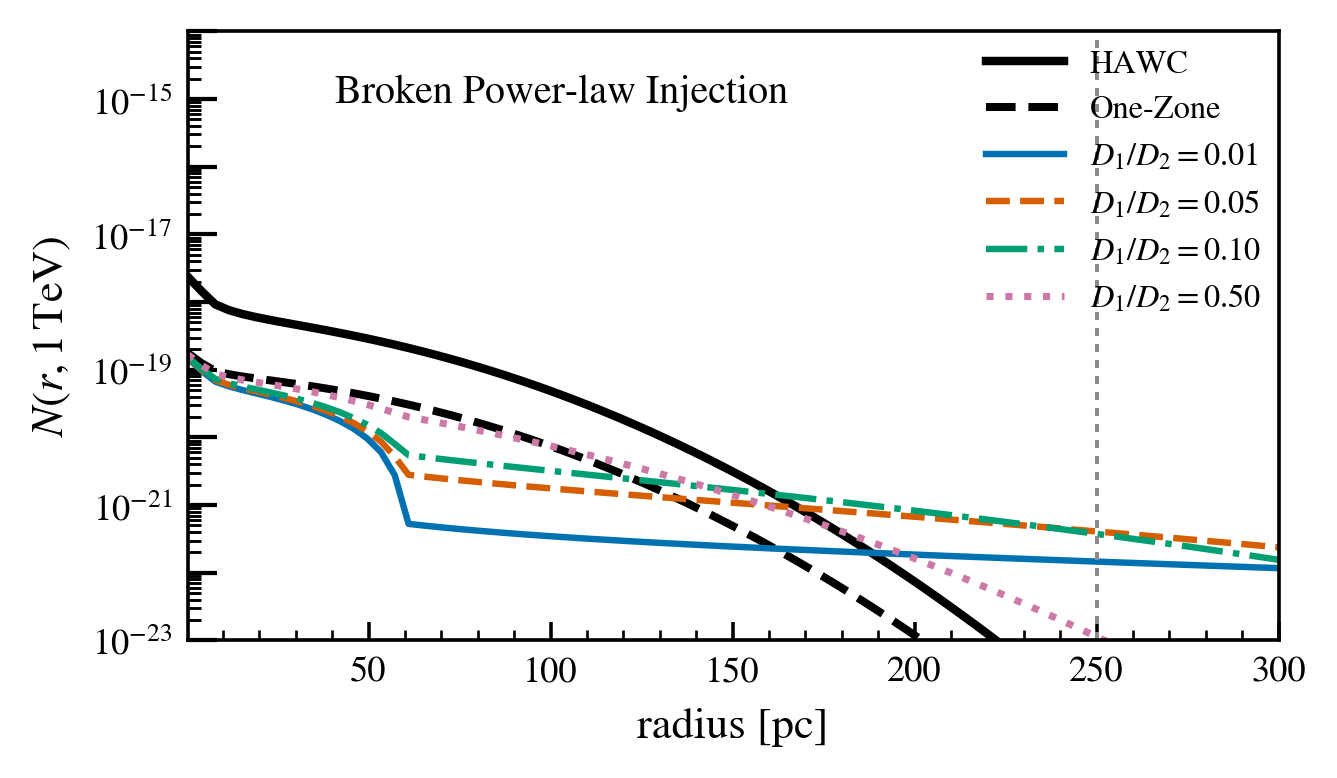}
\\
(b)
\end{minipage}

\caption{
$N(r,1\,\mathrm{TeV})$ for (a) single power-law injection and (b) broken power-law injection. The black solid line shows the HAWC benchmark, and the black dashed line denote the one-zone reference models, and the gray dashed line marks $r=250\,\mathrm{pc}$ .
}

\label{fig:density-spectrum}

\end{figure*}

Figure~\ref{fig:density-spectrum} shows $N(r,1\,\mathrm{TeV})$ for the single power-law and broken power-law injection models in panels (a) and (b), respectively. 
Both injection models show a clear change in slope at the boundary $r\simeq60\,\mathrm{pc}$. 
At distances of several hundred parsecs, the two-zone model gives a more extended tail than the one-zone model. 

\subsection{Gamma-ray surface brightness around Geminga}
\label{subsec:gamma_profile}

\begin{figure*}[htbp]

\centering

\begin{minipage}{0.48\textwidth}
\centering
\includegraphics[width=\linewidth]{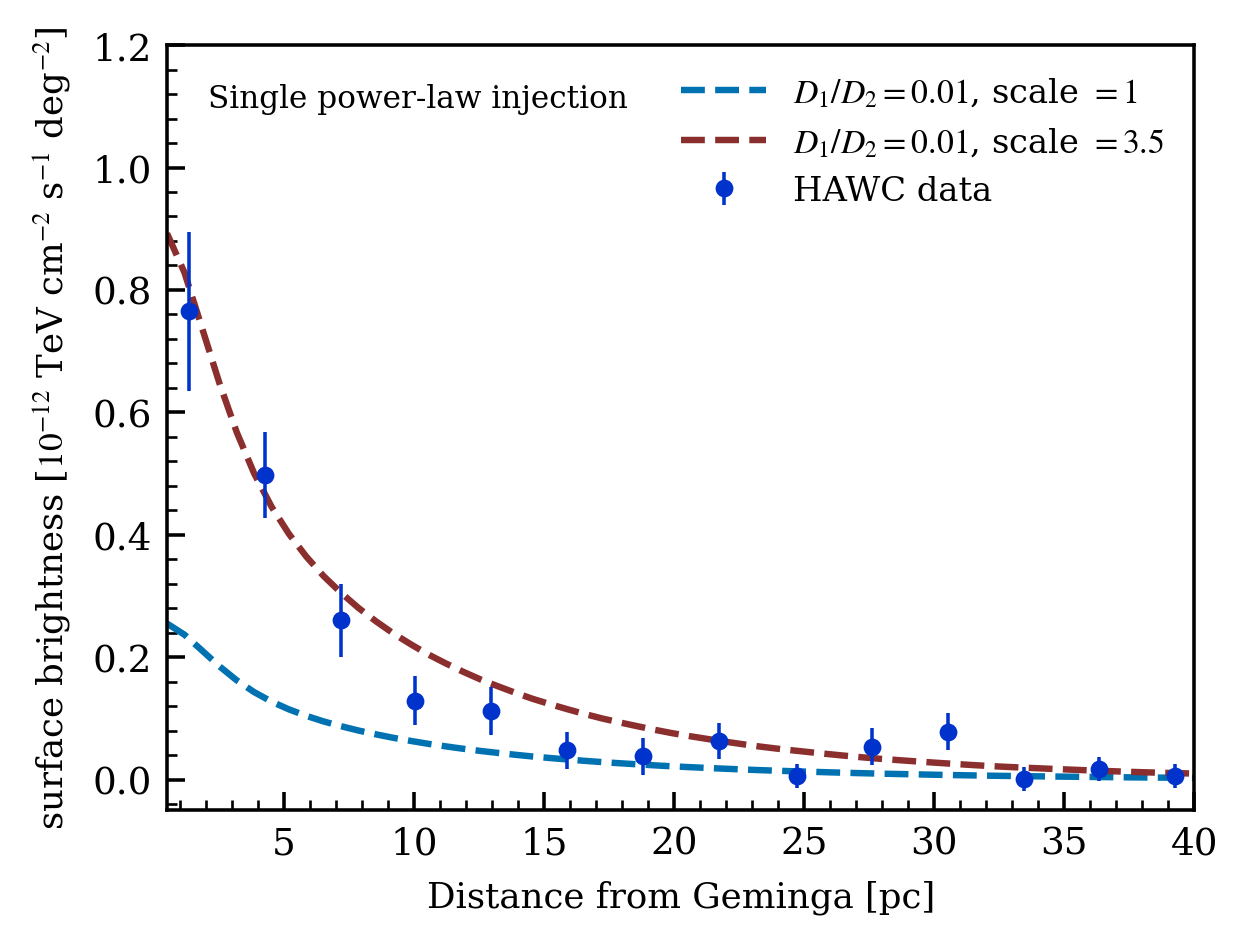}
\\
(a)
\end{minipage}
\hfill
\begin{minipage}{0.48\textwidth}
\centering
\includegraphics[width=\linewidth]{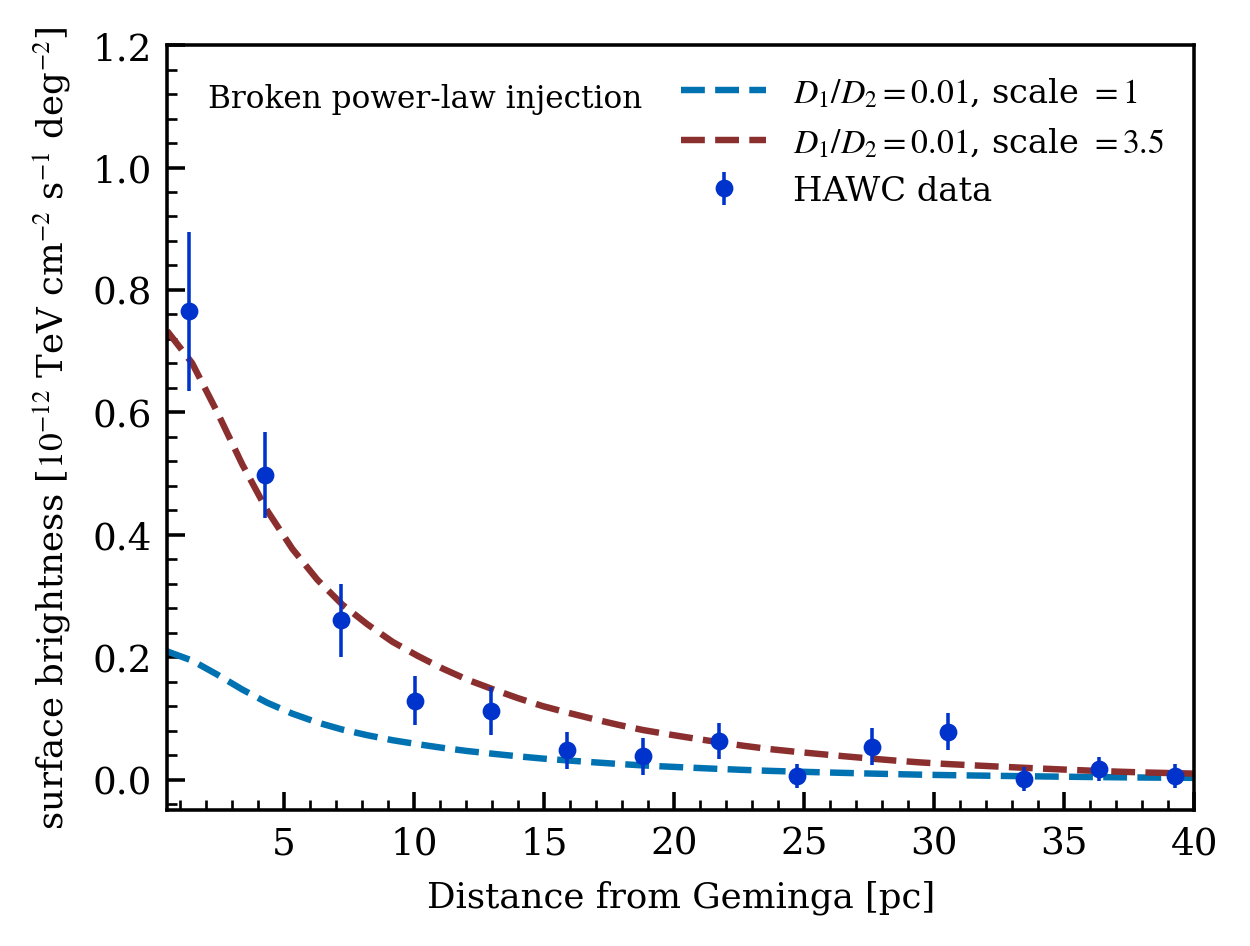}
\\
(b)
\end{minipage}

\caption{
Radial gamma-ray surface brightness around Geminga in the $5$--$50\,\mathrm{TeV}$ band for (a) single-power-law and (b) broken-power-law injection. Both panels use $D_1/D_2=0.01$. The blue dashed curves show the predictions with the efficiencies in Table~\ref{tab:injection_parameters}, red dashed show the same profiles multiplied by a factor of $3.5$, and the points are the HAWC measurements\citep{2017Sci...358..911A}.
}

\label{fig:hawc-profile}

\end{figure*}

Figure~\ref{fig:hawc-profile} compares the model predictions with the HAWC surface brightness data for the single power-law and broken power-law injection models. Both calculations use $D_1/D_2=0.01$, and the surface brightness is integrated over photon energies from $5-50\,\mathrm{TeV}$.
For $\mathrm{scale}=1$, the models show a radial decline similar to that of the HAWC data, but the predicted surface brightness is lower than the observations.
Tang \& Piran (2019) applied an overall normalization factor of $3.5$ when comparing their results with the HAWC surface-brightness data. They pointed out that this factor arises from differences in the gamma-ray flux normalization obtained by HAWC using different spatial templates, such as the disk and diffusion templates, rather than from the two-zone diffusion model itself.
Following this prescription, we multiply the model-predicted surface brightness by an overall factor of $\mathrm{scale}=3.5$. 

\subsection{Positron flux at the Earth}
\label{subsec:positron_flux}

\begin{figure*}[htbp]

\centering

\begin{minipage}{0.48\textwidth}
\centering
\includegraphics[width=\linewidth]{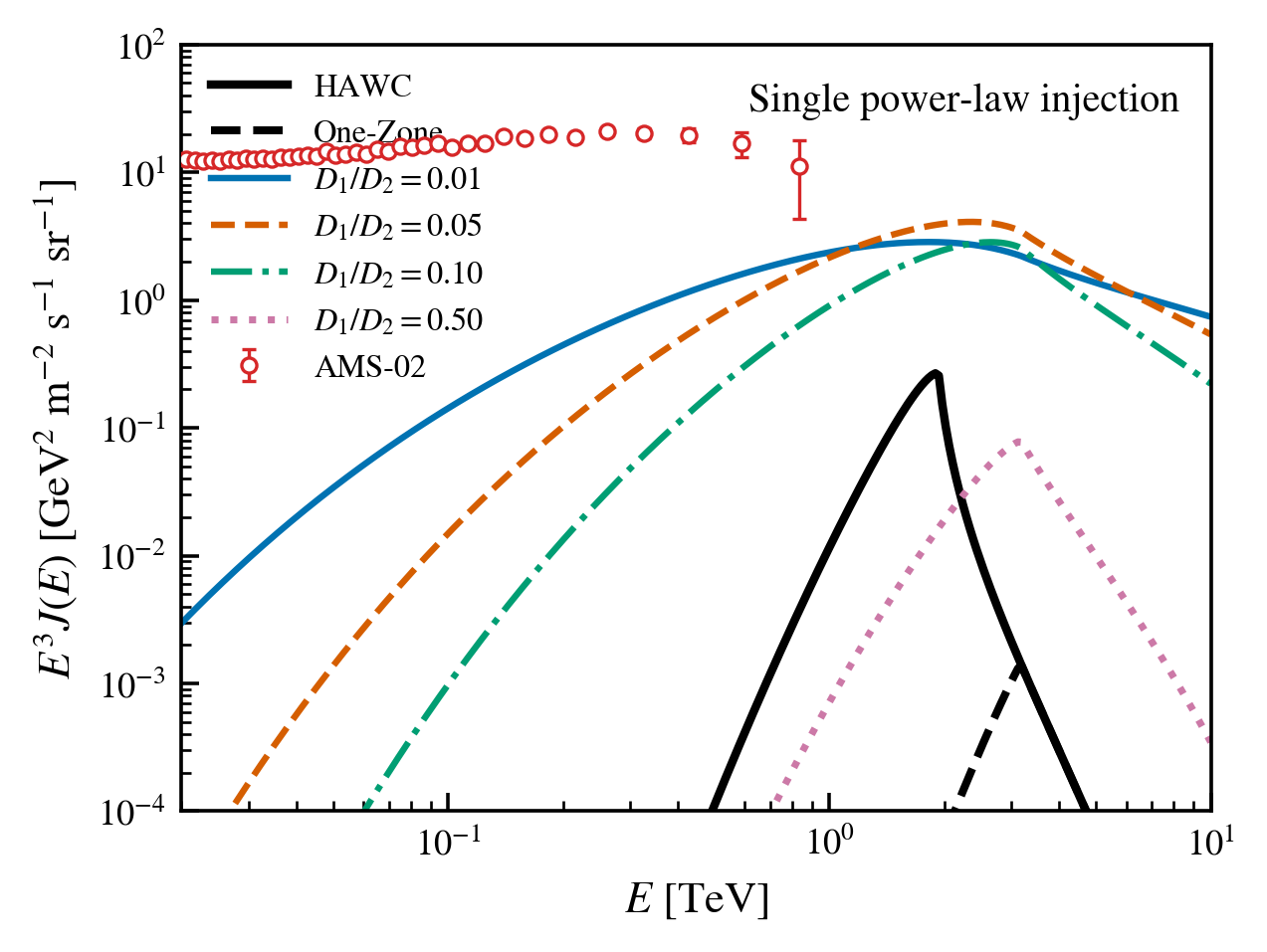}
\\
(a)
\end{minipage}
\hfill
\begin{minipage}{0.48\textwidth}
\centering
\includegraphics[width=\linewidth]{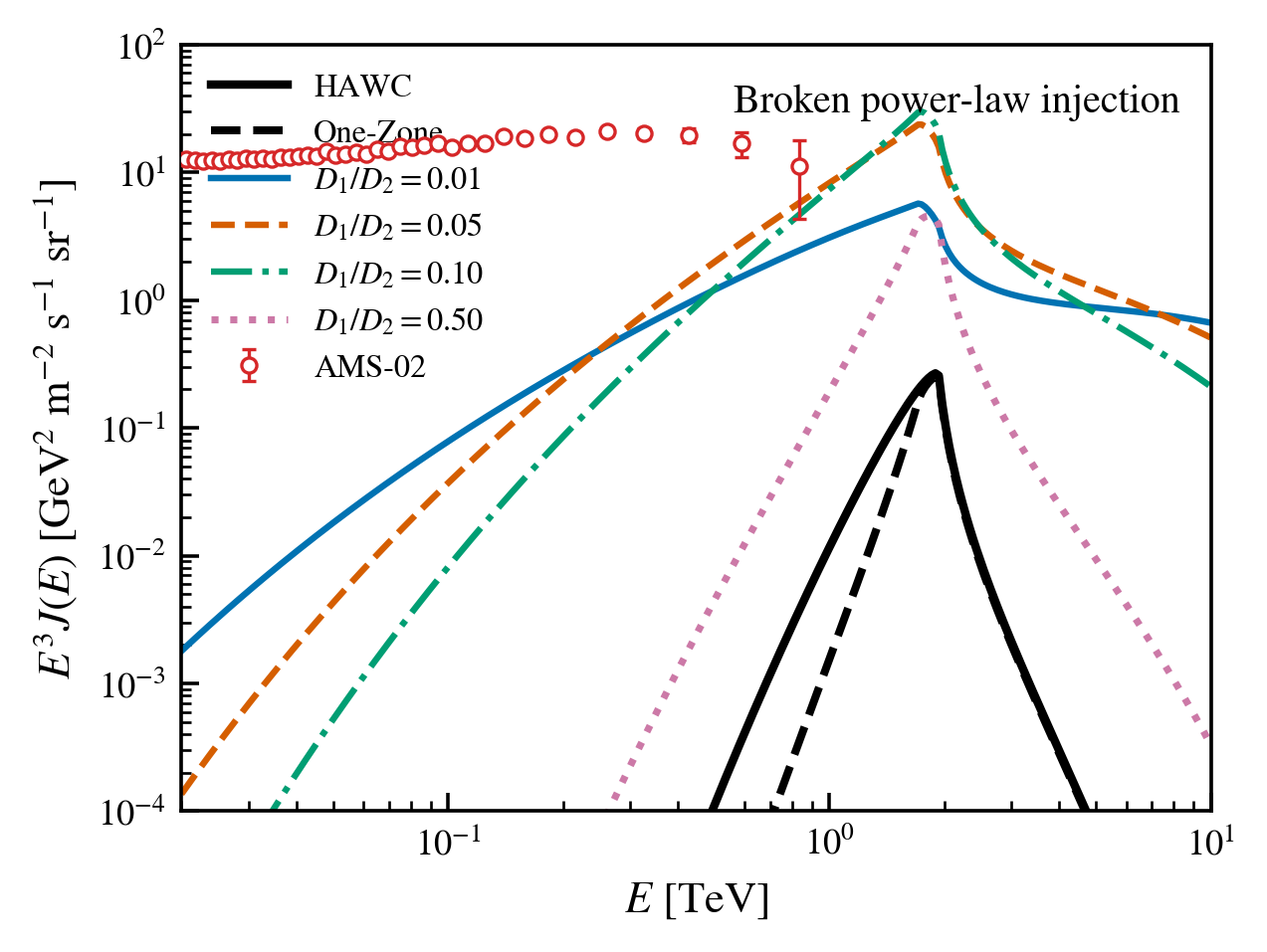}
\\
(b)
\end{minipage}

\caption{
Positron flux $E^3J(E)$ at the Earth for (a) single power-law and (b) broken power-law injection. The black solid and dashed curves denote the HAWC benchmark and the one-zone model, respectively, while the colored curves show two-zone models with different $D_1/D_2$. Parameters are listed in Tables~\ref{tab:propagation_parameters} and \ref{tab:injection_parameters}. The data are taken from AMS-02 \citep{PhysRevLett.122.101101}.
}

\label{fig:positron-flux}

\end{figure*}

We adopt a distance of $d=250\,\mathrm{pc}$ for Geminga when calculating the positron flux at the Earth. Figure~\ref{fig:positron-flux} shows $E^3J(E)$ for different values of $D_1/D_2$ for the single power-law and broken power-law injection models.

For single power-law injection, as shown in Fig.~\ref{fig:positron-flux}(a), the positron flux predicted by the one-zone model is confined to a relatively narrow energy range around the TeV scale and decreases rapidly at higher energies. For $D_1/D_2\leq0.1$, the two-zone models give higher fluxes over a broader energy range. Changing $D_1/D_2$ shifts the spectral peak and changes its height and width. 


For broken power-law injection, the two-zone models predict higher positron fluxes than the one-zone model in the TeV range, as shown in Fig.~\ref{fig:positron-flux}(b). The models with $D_1/D_2=0.05$ and $0.10$ exhibit the highest peaks, while the model with $D_1/D_2=0.01$ has a broader high-energy tail.

\begin{figure*}[htbp]
\centering
\begin{minipage}{0.48\textwidth}
\centering
\includegraphics[width=\linewidth]{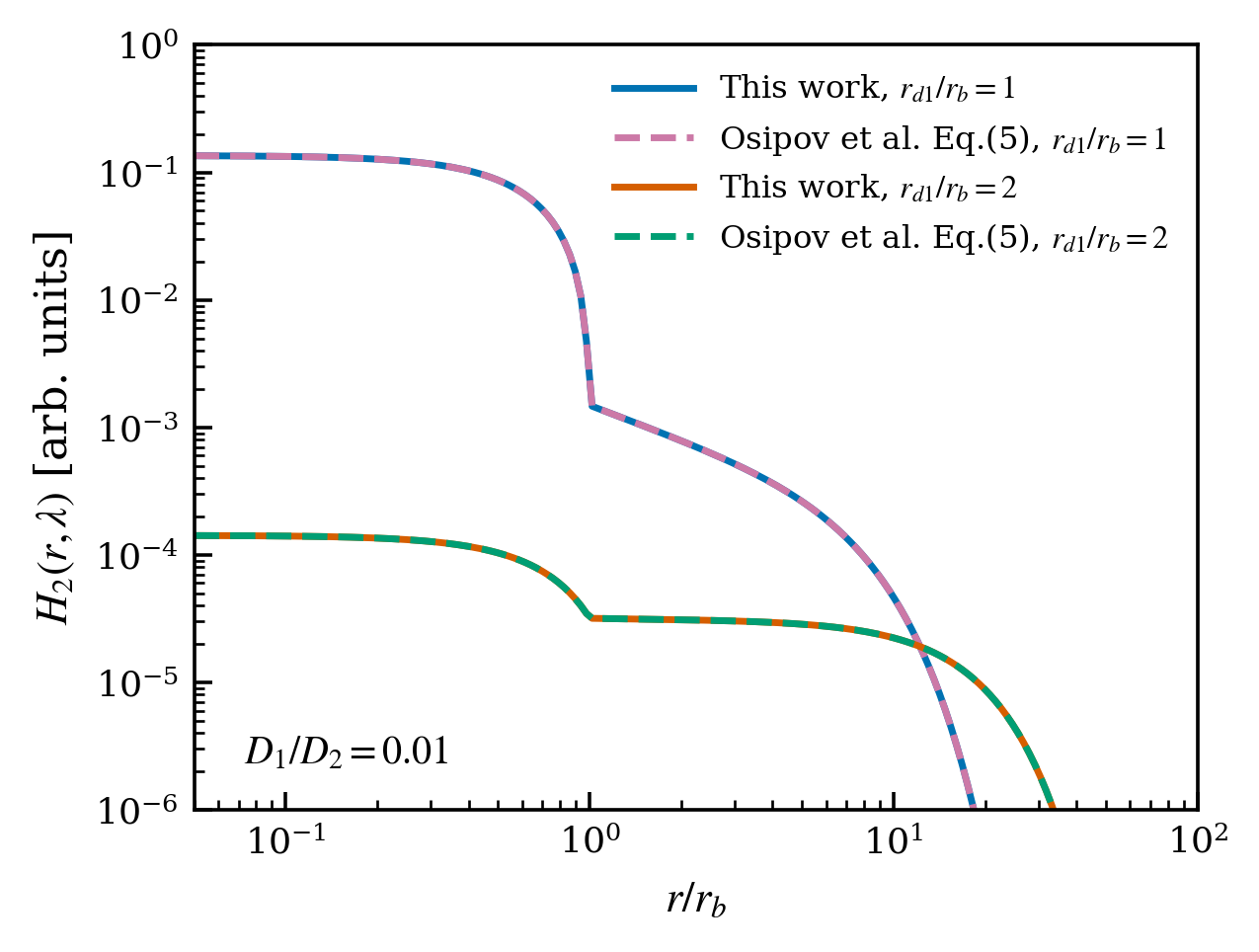}
\\
(a)
\end{minipage}
\hfill
\begin{minipage}{0.48\textwidth}
\centering
\includegraphics[width=\linewidth]{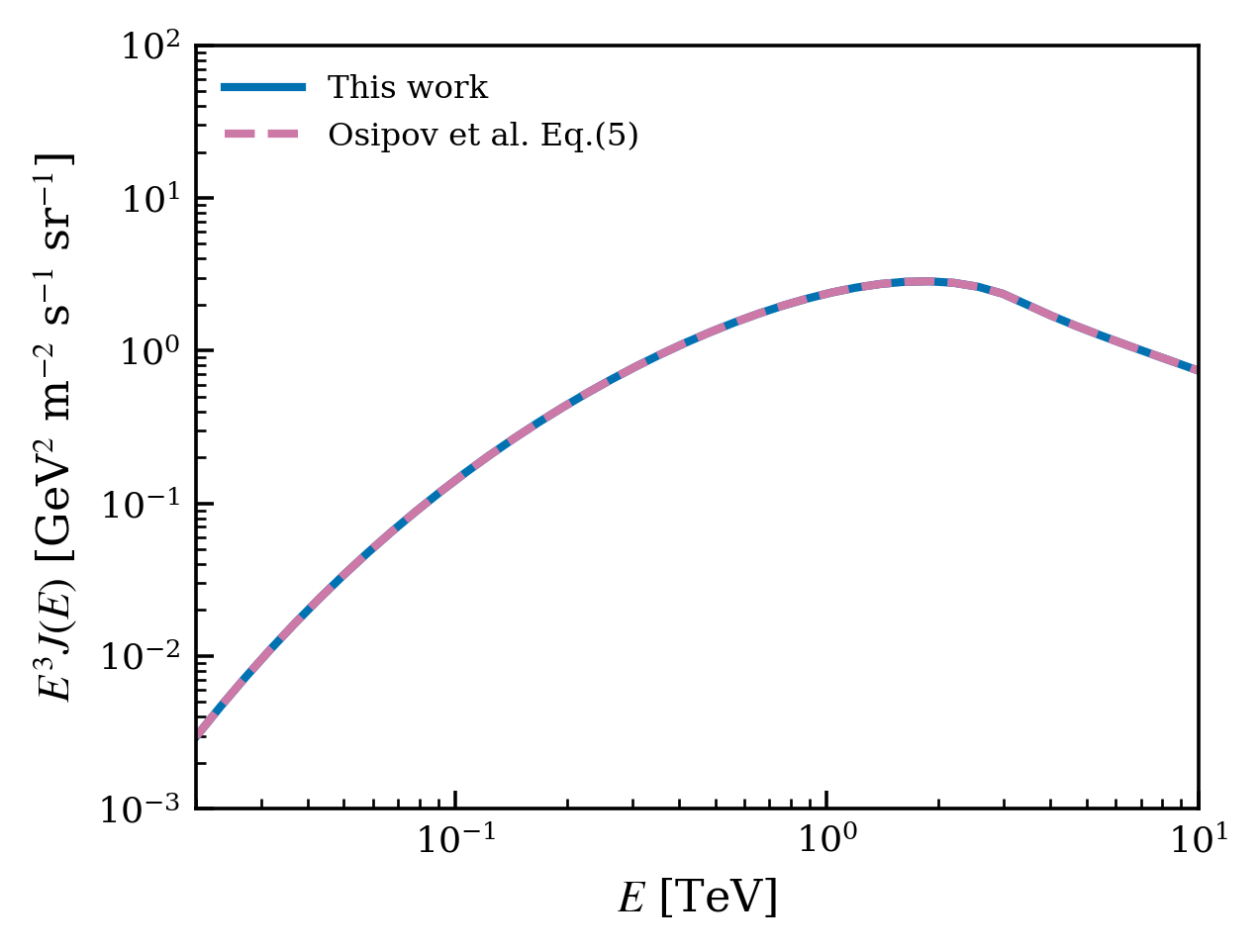}
\\
(b)
\end{minipage}
\caption{Comparison with the analytical results of \citet{2020JPhCS1697a2009O}. (a) Spatial diffusion kernel $H$. (b) Positron flux $E^3J(E)$ at the Earth. The same source parameters and $D_1/D_2=0.01$ are used in both calculations.}
\label{fig:comparison-osipov}
\end{figure*}

We further compare the spatial diffusion kernel and the resulting positron flux at the Earth obtained in this work with the analytical results of Osipov et al.~\citep{2020JPhCS1697a2009O}. 
Using the same source parameters and adopting $D_1/D_2=0.01$, the two analytical approaches give nearly identical results for the propagation kernel $H$ and the positron flux $E^3J(E)$, as shown in Fig.~\ref{fig:comparison-osipov}.



\section{Discussion and conclusions}
\label{sec:discussion}


We have derived an analytical Green's function for the two-zone diffusion problem using the Laplace transform.
The kernel derived here satisfies the continuity of the particle number density and diffusive flux at the slow-zone boundary $r_b$.
It also satisfies the particle-number normalization condition and reduces to the standard three-dimensional Gaussian kernel when $D_1=D_2$.
In addition, the propagation kernel and the resulting positron flux at the Earth agree with the independent integral representation of Osipov et al.~\citep{2020JPhCS1697a2009O}, providing a check of the series inversion and numerical summation used in this work.

We have examined how the slow-zone radius, the ratio of the inner and outer diffusion coefficients, and the injection spectrum affect the lepton distribution, the TeV gamma-ray surface brightness, and the positron flux.
A discontinuity in the diffusion coefficient produces a distinct change in slope near the boundary.
Relative to the one-zone solution, the two-zone propagation kernel develops a more extended outer tail.
The two-zone models also produce broader positron spectra and higher positron fluxes at the Earth than the one-zone models.
The injection spectrum further affects the height and shape of the high-energy peak in the positron spectrum.
When applied to the calculation of the TeV $\gamma$-ray surface brightness around Geminga, the propagation model captures the main spatial trend of the HAWC surface-brightness profile, namely its decrease with increasing projected distance. However, matching the absolute HAWC surface brightness requires an overall factor of 3.5, which may partly arise from uncertainties in the injected lepton spectrum and normalization, as well as from the spatial template adopted by HAWC.

The extended TeV emission detected by HAWC around Geminga and PSR B0656+14, together with the LHAASO detection of LHAASO J0621+3755 around PSR J0622+3749, suggest that pulsar halos may represent a broader class of structures associated with the escape and transport of high-energy lepton from pulsars\citep{2017Sci...358..911A,PhysRevLett.126.241103}.
The two-zone diffusion framework may also provide a useful approach for studying other pulsar halos in future work.

At the same time, suppressed diffusion is not the only possible explanation of the observed morphology. 
Recchia et al.~\citep{PhysRevD.104.123017} showed that a continuous transition from quasi-ballistic to diffusive propagation can reproduce the $\gamma$-ray observations of Geminga, Monogem, and PSR J0622+3749 using diffusion coefficients close to typical interstellar values. 

As HAWC, LHAASO, and other ground-based experiments continue to accumulate pulsar-halo observations with improved statistics, energy resolution, and angular resolution, two-zone diffusion models will be subject to more stringent tests.
Measurements of the halo size and surface-brightness profile in different energy bands can further constrain the slow-zone radius and the ratio of the diffusion coefficients between the inner and outer zones.
Systematic comparisons among different pulsar halos will also help determine whether suppressed diffusion near pulsars is a common phenomenon and distinguish two-zone diffusion from quasi-ballistic propagation and other particle-transport scenarios.

\section*{Acknowledgements}
This work is supported by the National Natural Science Foundation of China (Nos. 12333006, 12375107, 12575120). Y.S. acknowledges support from the National Natural Science Foundation of China (NSFC) with grant No. 12303001, and the Fundamental Research Funds for the Central Universities.

\bibliographystyle{aasjournal}
\bibliography{refs}

\appendix
\renewcommand{\theHequation}{\thesection.\arabic{equation}}
\section{Derivation of the Laplace-space Green's function}
\label{app:laplace_derivation}

In this Appendix, we derive the Laplace-space coefficients used in
Equations~(\ref{eq:H_laplace_inner_series}) and
(\ref{eq:H_laplace_outer_series}). 

We first solve the two-zone diffusion
problem in the radial coordinate $r$ and the diffusion-time variable $t$.
The Green's function $\mathcal{H}(r,t)$ for a unit instantaneous point source satisfies
\begin{equation}
\frac{\partial \mathcal{H}}{\partial t}
-
\frac{1}{r^2}
\frac{\partial}{\partial r}
\left[
r^2D_0(r)
\frac{\partial \mathcal{H}}{\partial r}
\right]
=
\frac{\delta(r)}{4\pi r^2}\delta(t),
\end{equation}

To streamline the notation, we define
\begin{equation}
D_1\equiv D_{1,0},
\qquad
D_2\equiv D_{2,0},
\qquad
k_i\equiv\sqrt{\frac{s}{D_i}},
\quad i=1,2,
\label{eq:app_def_ki}
\end{equation}
where $i=1$ denotes the inner slow-diffusion region and $i=2$
the outer region.

\subsection{General solutions and boundary conditions}
\label{app:general_solution}

We begin by introducing the Laplace transform with respect to $t$,
\begin{equation}
\widetilde{\mathcal{H}}(r,s)
=
\int_0^\infty
e^{-st}\mathcal{H}(r,t)\,dt,
\label{eq:app_H_laplace}
\end{equation}

After the transformation, the diffusion equation becomes

\begin{equation}
s\widetilde{\mathcal{H}}(r,s)
-
\frac{1}{r^2}
\frac{d}{dr}
\left[
r^2D_0(r)
\frac{d\widetilde{\mathcal{H}}}{dr}
\right]
=
\frac{\delta(r)}{4\pi r^2}.
\label{eq:app_laplace_full}
\end{equation}

The Laplace-space problem is most conveniently treated by separating three distinct cases. First, away from both the point source and the diffusion boundary, $r\neq0,r_b$, the equation is homogeneous and $D_0(r)$ is constant within each region. 
Second, the point-source condition must be imposed at $r=0$. Third, the two regional solutions must be matched at $r=r_b$.

For $r\neq0,r_b$, Equation~(\ref{eq:app_laplace_full}) reduces in each region to
\begin{equation}
s\widetilde{\mathcal H}_i
-
\frac{D_i}{r^2}
\frac{d}{dr}
\left(
r^2\frac{d\widetilde{\mathcal H}_i}{dr}
\right)
=0,
\label{eq:app_radial_equation}
\end{equation}

Since the diffusion coefficient is constant within each region,
$\partial_r D_i=0$, the radial operator can be simplified using
\begin{equation}
\frac{1}{r^2}
\frac{d}{dr}
\left(
r^2\frac{d\widetilde{\mathcal H}_i}{dr}
\right)
=
\frac{1}{r}
\frac{d^2}{dr^2}
\left(
r\widetilde{\mathcal H}_i
\right).
\label{eq:app_radial_identity}
\end{equation}
Therefore, for $r\neq0,r_b$,
Equation~(\ref{eq:app_radial_equation}) becomes
\begin{equation}
s\widetilde{\mathcal H}_i
-
\frac{D_i}{r}
\frac{d^2}{dr^2}
\left(
r\widetilde{\mathcal H}_i
\right)
=0,
\qquad i=1,2.
\label{eq:app_radial_simplified}
\end{equation}

Defining $u_i(r,s) \equiv
r\widetilde{\mathcal H}_i(r,s)$, and using $k_i^2=s/D_i$, we obtain

\begin{equation}
\frac{d^2u_i}{dr^2}
-
k_i^2u_i
=0.
\label{eq:app_u_equation}
\end{equation}
The general solutions in the two regions are therefore
\begin{equation}
\widetilde{\mathcal H}_{\rm in}(r,s)
=
\frac{
A e^{-k_1r}
+
B e^{k_1r}
}{r},
\qquad
0<r<r_b,
\label{eq:app_inner_general}
\end{equation}
and
\begin{equation}
\widetilde{\mathcal H}_{\rm out}(r,s)
=
\frac{C e^{-k_2r}}{r},
\qquad
r>r_b.
\label{eq:app_outer_general_full}
\end{equation}
where the exponentially growing outer solution has been discarded by the condition $\widetilde{\mathcal H}_{\rm out}\rightarrow0$ as $r\rightarrow\infty$.

The source position $r=0$ and the boundary $r=r_b$  should be treated directly using the full Laplace-space equation, Equation~(\ref{eq:app_laplace_full}). Integrating Equation~(\ref{eq:app_laplace_full}) over a small sphere $0<r<\epsilon$ gives
\begin{equation}
\int_0^\epsilon
\left[
s\widetilde{\mathcal H}_{\rm in}
-
\frac{1}{r^2}
\frac{d}{dr}
\left(
r^2D_1
\frac{d\widetilde{\mathcal H}_{\rm in}}{dr}
\right)
\right]
4\pi r^2\,dr
=
\int_0^\epsilon
\frac{\delta(r)}{4\pi r^2}
\,4\pi r^2\,dr .
\label{eq:app_origin_sphere_integral}
\end{equation}
Since the sphere contains the point source, the right-hand side is
equal to unity. Therefore,
\begin{equation}
4\pi s
\int_0^\epsilon
\widetilde{\mathcal H}_{\rm in}(r,s)\,r^2\,dr
-
4\pi D_1
\left[
r^2
\frac{d\widetilde{\mathcal H}_{\rm in}}{dr}
\right]_{0}^{\epsilon}
=
1.
\label{eq:app_origin_integrated}
\end{equation}

In the limit $\epsilon\rightarrow0$, the volume term vanishes, and the
point-source normalization becomes
\begin{equation}
-4\pi D_1
\lim_{r\rightarrow0^+}
r^2
\frac{d\widetilde{\mathcal H}_{\rm in}}{dr}
=
1.
\label{eq:app_origin_flux}
\end{equation}
which gives 
\begin{equation}
A+B
=
\frac{1}{4\pi D_1}.
\label{eq:app_origin_condition}
\end{equation}

At the boundary $r=r_b$, both the particle density and the diffusive flux are continuous
\begin{equation}
\widetilde{\mathcal H}_{\rm in}(r_b,s)
=
\widetilde{\mathcal H}_{\rm out}(r_b,s),
\label{eq:app_interface_density}
\end{equation}
and
\begin{equation}
D_1
\left.
\frac{d\widetilde{\mathcal H}_{\rm in}}{dr}
\right|_{r=r_b}
=
D_2
\left.
\frac{d\widetilde{\mathcal H}_{\rm out}}{dr}
\right|_{r=r_b}.
\label{eq:app_interface_flux}
\end{equation}

Solving the three conditions above determines the coefficients $A$, $B$, and $C$.

\subsection{Laplace-space coefficients}
\label{app:laplace_coefficients}

Solving Equations~(\ref{eq:app_origin_condition})--(\ref{eq:app_interface_flux}) gives
\begin{align}
A
&=
\frac{1}{4\pi D_1}
\frac{
D_1(1-k_1r_b)-D_2(1+k_2r_b)
}{D_1
\left[
(1-k_1r_b)-e^{-2k_1r_b}(1+k_1r_b)
\right]
-
D_2(1+k_2r_b) \left(1 - e^{-2k_1 r_b}\right)},
\label{eq:app_coeff_A}
\\
B
&=
\frac{1}{4\pi D_1}
\frac{
e^{-2k_1r_b}
\left[-D_1(1+k_1r_b)+D_2(1+k_2r_b)\right]
}{D_1
\left[
(1-k_1r_b)-e^{-2k_1r_b}(1+k_1r_b)
\right]
-
D_2(1+k_2r_b)\left(1 - e^{-2k_1 r_b}\right)},
\label{eq:app_coeff_B}
\\
C
&=
-\frac{1}{4\pi}
\frac{
2k_1r_b\,e^{(k_2-k_1)r_b}
}{D_1
\left[
(1-k_1r_b)-e^{-2k_1r_b}(1+k_1r_b)
\right]
-
D_2(1+k_2r_b) \left(1 - e^{-2k_1 r_b}\right)},
\label{eq:app_coeff_C}
\end{align}

For later convenience, we define
\begin{align}
M&\equiv D_1(1-k_1r_b)-D_2(1+k_2r_b),
\label{eq:app_M}
\\
N&\equiv -D_1(1+k_1r_b)+D_2(1+k_2r_b),
\label{eq:app_N}
\end{align}
so that
\begin{equation}
D_1
\left[
(1-k_1r_b)-e^{-2k_1r_b}(1+k_1r_b)
\right]
-
D_2(1+k_2r_b) \left(1 - e^{-2k_1 r_b}\right)=M+N e^{-2k_1r_b}.
\label{eq:app_delta_MN}
\end{equation}
The two regional solutions can then be written in the compact form
\begin{align}
\widetilde{\mathcal H}_{\rm in}(r,s)
&=
\frac{1}{4\pi rD_1}
\frac{
M e^{-k_1r}
+
N e^{-k_1(2r_b-r)}
}{
M+N e^{-2k_1r_b}
},
\label{eq:app_inner_compact}
\\
\widetilde{\mathcal H}_{\rm out}(r,s)
&=
-\frac{1}{4\pi r}
\frac{
2k_1r_b\,e^{-k_1r_b}e^{-k_2(r-r_b)}
}{
M+N e^{-2k_1r_b}
}.
\label{eq:app_outer_compact}
\end{align}

\subsection{Series representation}
\label{app:series_representation}

The common denominator can be expanded as
\begin{equation}
\frac{1}{M+N e^{-2k_1r_b}}
=
\frac{1}{M}
\sum_{n=0}^{\infty}
(-1)^n
\left(\frac{N}{M}\right)^n
 e^{-2nk_1r_b},
\label{eq:app_geometric_series}
\end{equation}
provided that
\begin{equation}
\left|
\frac{N}{M}e^{-2k_1r_b}
\right|<1 .
\label{eq:app_series_condition}
\end{equation}

Equations~(\ref{eq:app_inner_compact}) and
(\ref{eq:app_outer_compact}) can be written as
\begin{align}
\widetilde{\mathcal H}_{\rm in}(r,s)
={}&
\frac{1}{4\pi rD_1}
\sum_{n=0}^{\infty}(-1)^n
\Bigg[
\left(\frac{N}{M}\right)^n
 e^{-k_1(2nr_b+r)}
+
\left(\frac{N}{M}\right)^{n+1}
e^{-k_1[2(n+1)r_b-r]}
\Bigg],
\label{eq:app_inner_NM_series}
\\
\widetilde{\mathcal H}_{\rm out}(r,s)
={}&
-\frac{1}{4\pi r}
\sum_{n=0}^{\infty}(-1)^n
\left(\frac{N}{M}\right)^n
\frac{2k_1r_b}{M}
\,
 e^{-\left[
k_1{(2n+1)r_b}
+
k_2{(r-r_b)}
\right]}.
\label{eq:app_outer_NM_series}
\end{align}

Using $k_2=\sqrt{D_1/D_2}\,k_1$, the ratio $N/M$ can be written as
\begin{equation}
\frac{N}{M}
=
R_{\rm c}
+
\frac{R_0}{\sqrt{s}+\lambda},
\label{eq:app_N_over_M}
\end{equation}
where
$R_{\rm c}=(\sqrt{D_1}-\sqrt{D_2})/(\sqrt{D_1}+\sqrt{D_2})$,
$R_0=2(D_1-D_2)\sqrt{D_1}/[r_b(\sqrt{D_1}+\sqrt{D_2})^2]$, and
$\lambda=-(D_1-D_2)/[r_b(\sqrt{D_1}+\sqrt{D_2})]$.

Consequently,
\begin{equation}
\left(\frac{N}{M}\right)^n
=
\sum_{m=0}^{n}
\binom{n}{m}
R_{\rm c}^{\,n-m}R_0^{\,m}
\frac{1}{(\sqrt{s}+\lambda)^m}.
\label{eq:app_binomial_expansion}
\end{equation}

And $2k_1r_b/M$ in the outer solution can be written as
\begin{equation}
\frac{2k_1r_b}{M}
=
2r_b
\left[
\frac{1}{\delta_0}
-
\frac{\lambda}{\delta_0}
\frac{1}{\sqrt{s}+\lambda}
\right],
\qquad
\delta_0
\equiv
-r_b\left(D_1+\sqrt{D_1D_2}\right).
\label{eq:app_outer_prefactor}
\end{equation}

Substituting Equations~(\ref{eq:app_binomial_expansion}) and
(\ref{eq:app_outer_prefactor}) into
Equations~(\ref{eq:app_inner_NM_series}) and
(\ref{eq:app_outer_NM_series}) we obtain

\begin{equation}
\begin{aligned}
\widetilde{\mathcal{H}}_{\mathrm{in}}(r,s)
={}&
\frac{1}{4\pi rD_1}
\sum_{n=0}^{\infty}
(-1)^n
\Bigg[
\\
&\sum_{m=0}^{n}
\binom{n}{m}
R_{\mathrm{c}}^{\,n-m}
R_0^{\,m}
\frac{1
}{
\left(\sqrt{s}+\lambda\right)^m
}
\exp\left[
-\sqrt{s}
\frac{2nr_b+r}{\sqrt{D_1}}
\right]
\\
&+
\sum_{m=0}^{n+1}
\binom{n+1}{m}
R_{\mathrm{c}}^{\,n+1-m}
R_0^{\,m}
\frac{1
}{
\left(\sqrt{s}+\lambda\right)^m
}
\exp\left[
-\sqrt{s}
\frac{2(n+1)r_b-r}{\sqrt{D_1}}
\right]
\Bigg].
\end{aligned}
\label{eq:H_laplace_inner_}
\end{equation}

\begin{equation}
\begin{aligned}
\widetilde{\mathcal{H}}_{\mathrm{out}}(r,s)
={}&
-\frac{2r_b}{4\pi r}
\sum_{n=0}^{\infty}
(-1)^n
\Bigg\{
\\
&\frac{1}{-r_b
\left(
D_{1}
+
\sqrt{D_{1}D_{2}}
\right)}
\sum_{m=0}^{n}
\binom{n}{m}
R_{\mathrm{c}}^{\,n-m}
R_0^{\,m}
\frac{1}{
\left(\sqrt{s}+\lambda\right)^m
}
\\
&-
\frac{\lambda}{-r_b
\left(
D_{1}
+
\sqrt{D_{1}D_{2}}
\right)}
\sum_{m=0}^{n}
\binom{n}{m}
R_{\mathrm{c}}^{\,n-m}
R_0^{\,m}
\frac{1}{
\left(\sqrt{s}+\lambda\right)^{m+1}
}
\Bigg\}
\\
&\times
\exp
\left\{
-\sqrt{s}
\left[
\frac{(2n+1)r_b}{\sqrt{D_{1}}}
+
\frac{r-r_b}{\sqrt{D_{2}}}
\right]
\right\}.
\end{aligned}
\label{eq:H_laplace_outer_}
\end{equation}

\subsection{Term-by-term inverse Laplace transformation}
\label{app:inverse_laplace}

Each term in the resulting series contains a factor of the generic form
\begin{equation}
\frac{e^{-a\sqrt{s}}}{(\sqrt{s}+\lambda)^m}.
\label{eq:app_generic_term}
\end{equation}

For $m=1$, the inverse transform is
\begin{equation}
\begin{aligned}
f_1(a,\lambda,t)
&\equiv
\mathcal L^{-1}
\left\{
\frac{e^{-a\sqrt{s}}}{\sqrt{s}+\lambda}
\right\}
=
\frac{1}{\sqrt{\pi t}}
\exp\left(-\frac{a^2}{4t}\right)
-
\lambda e^{a\lambda+\lambda^2t}
\operatorname{erfc}
\left(
\frac{a}{2\sqrt{t}}+\lambda\sqrt{t}
\right).
\end{aligned}
\label{eq:app_inverse_m1}
\end{equation}

For integer $m\geq1$,
\begin{equation}
\begin{aligned}
f_m(a,\lambda,t)
&\equiv
\mathcal L^{-1}
\left\{
\frac{e^{-a\sqrt{s}}}{(\sqrt{s}+\lambda)^m}
\right\}
=
\frac{(-1)^{m-1}}{(m-1)!}
\frac{\partial^{m-1}}{\partial\lambda^{m-1}}
\mathcal L^{-1}
\left\{
\frac{e^{-a\sqrt{s}}}{\sqrt{s}+\lambda}
\right\},    
\end{aligned}
\label{eq:app_inverse_general_m}
\end{equation}

whereas the $m=0$ term is
\begin{equation}
\begin{aligned}
f_0(a,t)
\equiv
\mathcal L^{-1}
\left\{e^{-a\sqrt{s}}\right\}
=
\frac{a}{2\sqrt{\pi}\,t^{3/2}}
\exp\left(-\frac{a^2}{4t}\right).
\end{aligned}
\label{eq:app_inverse_m0}
\end{equation}

\subsection{Green's function in the radial diffusion-time space}
\label{app:rt_green_function}

Using the inverse Laplace transforms defined above, the Green's
function can be written as
\begin{equation}
\mathcal H(r,t)
=
\begin{cases}
\mathcal H_{\rm in}(r,t), & 0<r<r_b,\\[4pt]
\mathcal H_{\rm out}(r,t), & r\geq r_b.
\end{cases}
\label{eq:app_H_rt_piecewise}
\end{equation}

For convenience, we introduce
\begin{equation}
a_n^{(+)}
\equiv
\frac{2nr_b+r}{\sqrt{D_1}},
\qquad
a_n^{(-)}
\equiv
\frac{2(n+1)r_b-r}{\sqrt{D_1}},
\label{eq:app_inner_path_lengths}
\end{equation}
and
\begin{equation}
a_n^{\rm out}
\equiv
\frac{(2n+1)r_b}{\sqrt{D_1}}
+
\frac{r-r_b}{\sqrt{D_2}}.
\label{eq:app_outer_path_length}
\end{equation}

The inner-region solution is then
\begin{align}
\mathcal H_{\rm in}(r,t)
={}&
\frac{1}{4\pi rD_1}
\sum_{n=0}^{\infty}
(-1)^n
\Bigg[
\nonumber\\
&
\sum_{m=0}^{n}
\binom{n}{m}
R_{\rm c}^{\,n-m}
R_0^{\,m}
f_m
\left(
a_n^{(+)},
\lambda,
t
\right)
\nonumber\\
&+
\sum_{m=0}^{n+1}
\binom{n+1}{m}
R_{\rm c}^{\,n+1-m}
R_0^{\,m}
f_m
\left(
a_n^{(-)},
\lambda,
t
\right)
\Bigg],
\qquad 0<r<r_b.
\label{eq:app_H_rt_inner}
\end{align}

For the outer-region solution, the result is
\begin{align}
\mathcal H_{\rm out}(r,t)
={}&
-\frac{2r_b}{4\pi r}
\sum_{n=0}^{\infty}
(-1)^n
\Bigg[
\nonumber\\
&
\frac{1}{-r_b
\left(
D_1+\sqrt{D_1D_2}\right)}
\sum_{m=0}^{n}
\binom{n}{m}
R_{\rm c}^{\,n-m}
R_0^{\,m}
f_m
\left(
a_n^{\rm out},
\lambda,
t
\right)
\nonumber\\
&-
\frac{\lambda}{-r_b\left(
D_1+\sqrt{D_1D_2}\right)}
\sum_{m=0}^{n}
\binom{n}{m}
R_{\rm c}^{\,n-m}
R_0^{\,m}
f_{m+1}
\left(
a_n^{\rm out},
\lambda,
t
\right)
\Bigg],
\qquad r\geq r_b.
\label{eq:app_H_rt_outer}
\end{align}

\subsection{Including continuous energy losses}
\label{app:energy_mapping}

To include continuous energy losses for electrons and positrons, we assume
\begin{equation}
D(E,r)=\kappa(E)D_0(r)
\label{eq:app_D_factorized}
\end{equation}
we define 
\begin{equation}
u(E,E_0)
\equiv
\int_E^{E_0}
\frac{\kappa(E')}{b(E')}\,dE'.
\label{eq:app_common_pseudotime}
\end{equation}

For a particle injected with energy $E_0$ and observed at energy $E$, the
cooling time is
\begin{equation}
\tau(E,E_0)
=
\int_E^{E_0}
\frac{dE'}{b(E')} .
\label{eq:app_cooling_time}
\end{equation}

The corresponding cooling-dependent diffusion length in the $i$th diffusion
zone is defined as
\begin{equation}
r_{d,i}(E,E_0)
=
2
\left[
\int_E^{E_0}
\frac{D_i(E')}{b(E')}dE'
\right]^{1/2},
\qquad i=1,2 .
\label{eq:app_rd_def}
\end{equation}

Equivalently, 
\begin{equation}
\Lambda_i(E,E_0)
=
\int_E^{E_0}
\frac{D_i(E')}{b(E')}dE',
\label{eq:app_Lambda_i}
\end{equation}

It then follows that
\begin{equation}
\Lambda_i(E,E_0)
=
D_{i,0}\,u(E,E_0),
\qquad i=1,2,
\label{eq:app_Lambda_factorized}
\end{equation}
so that the replacement
\begin{equation}
D_{i,0}t
\longrightarrow
\Lambda_i(E,E_0)
\label{eq:app_Dt_replacement}
\end{equation}
and
\begin{equation}
r_{d,i}(E,E_0)
=2\sqrt{\Lambda_i(E,E_0)}
=2\sqrt{D_{i,0}u(E,E_0)}.
\label{eq:app_rd_mapping}
\end{equation}

Lastly, we obtain $H(r;E,E_0)$ given in
Equations~(\ref{eq:H_inner_expanded_q}) and
(\ref{eq:H_outer_expanded_q}).

\end{document}